\documentclass[11pt]{article}
\pdfoutput=1
\usepackage{geometry}                
\usepackage{graphicx}
\usepackage{amsmath}
\usepackage{amssymb}
\usepackage{authblk}
\usepackage[labelformat=simple]{subfig}

\usepackage{subfloat}
\usepackage{float}
\usepackage[scr=boondoxo,scrscaled=1.05]{mathalfa}

\usepackage{epstopdf}
\usepackage[backend=bibtex,style=numeric,sorting=none]{biblatex}
\renewbibmacro{in:}{}
\DeclareFieldFormat[article]{volume}{\mkbibbold{#1}}
\ifdefined\blackandwhite
\usepackage[pdftex]{hyperref} 
\else
\usepackage[colorlinks,pdftex]{hyperref} 
\hypersetup{citecolor = blue, linkcolor = red} 
\fi

\title{An Efficient Routing Protocol for Quantum Key Distribution Networks}
\author{Jia-Meng Yao $^{1,\dagger}$, Ya-Xing Wang $^{1,\dagger}$, Qiong Li $^{1,}$*, Hao-Kun Mao $^{1}$, Ahmed A.Abd El-Latif $^{2,3}$ and Nan Chen $^{4}$}
\affil{$^{1}${Department of Computer Science and Technology, Harbin Institute of Technology, Harbin 150080, China; qiongli@hit.edu.cn (Q.L.); 20B903022@stu.hit.edu.cn (J.Y.); yaxwang@163.com (Y.W.); icthkmao@gmail.com (H.M.)}\\$^{2}${EIAS Data Science Lab, College of Computer and Information Sciences, Prince Sultan University,Riyadh 11586, Saudi Arabia; aabdellatif@psu.edu.sa or a.rahiem@gmail.com}\\$^{3}${Department of Mathematics and Computer Science, Faculty of Science, Menoufia University,Shibin El Kom}\\$^{4}${School of Foreign Languages, Harbin Institute of Technology, Harbin 150080, China; chennan2010@hit.edu.cn}}

\date{}    

\begin{document}

\maketitle

\begin{abstract}
Quantum key distribution (QKD) can provide point-to-point information-theoretic secure key services for two connected users. In fact, the development of QKD networks needs more focus from the scientific community in order to broaden the service scale of QKD technology to deliver end-to-end secure key services. Of course, some recent efforts have been made to develop secure communication protocols based on QKD. However, due to the limited key generation capability of QKD devices, high quantum secure key utilization is the major concern for QKD networks. Since traditional routing techniques do not account for the state of quantum secure keys on links, applying them in QKD networks directly will result in underutilization of quantum secure keys. Therefore, an~efficient routing protocol for QKD networks, especially for large-scale QKD networks, is desperately needed. In this study, an efficient routing protocol based on optimized link-state routing, namely QOLSR, is proposed for QKD networks. QOLSR considerably improves quantum key utilization in QKD networks through link-state awareness and path optimization. Simulation results demonstrate the validity and efficiency of the proposed QOLSR routing protocol. Most importantly, with the growth of communication traffic, the benefit becomes even more apparent.
\end{abstract}

\section{Introduction}
The risks and challenges threatening network security have gradually increased with the rapid development of networks in recent years. It is therefore crucial to ensure the security of transmitted information in the network. Quantum communication is a novel secure communication technology based on the fundamental concepts and properties of quantum mechanics~\cite{1,2,3,4}. Quantum secure communication based on QKD is the most-rapidly evolving quantum communication technology currently in use. QKD is a~key-distribution technology in which the two parties of the communication first encode and transmit the key information using the quantum state as the information carrier, and~then negotiate the key between them. Quantum states can represent a wide range of binary data combinations due to their superposition. {Quantum key generation is typically performed by QKD using protocols such as BB84~\cite{5}, twin-field (TF) \cite{6,7}, and~so on~\cite{8,9}. }The security of QKD is based on the Heisenberg uncertainty principle and the quantum no-cloning theorem in quantum mechanics~\cite{2,3}. The~combination of QKD and a one-time pad (OTP) algorithm has the unique advantage of information-theoretical security~\cite{10,11}. However, QKD can only provide point-to-point information-theoretic secure keys for two connected users. To~broaden the service scale of QKD technology, study of QKD networks that can deliver end-to-end secure key services for more users is crucial~\cite{12,13,14}. Thanks to years of intense research on QKD networks, the~number of nodes in experimental QKD networks has expanded from 6 to 56, and~the transmission distance has increased from 19.6 km to 7600 km~\cite{15,16,17}.

To maintain the security of data packets transmitted in the QKD network, they must be encrypted with quantum secure keys. In~this study, it is assumed that routing packets must also be encrypted with quantum secure keys in order to guarantee the security of routing data. Therefore, QKD networks are characterized by their reliance on quantum secure keys in communication. Since traditional routing protocols do not account for this characteristic of QKD networks, applying them in QKD networks directly will result in the underutilization of quantum secure keys. That is not desirable because quantum secure keys are very precious in QKD networks. Consequently, a~routing protocol that can achieve high quantum secure key utilization is crucial to QKD~networks.

Many scholars have looked at routing protocols in QKD networks in order to address the routing problem. For~example, the~open shortest path first (OSPF) protocol has been used in the experimental DARPA QKD network~\cite{18}. Later, the~upgraded version, OSPF-v2 protocol, was used in the SECOQC network~\cite{19}. The~authors of~\cite{19} used the number of quantum secure keys  
instead of hop count as the routing metric for local load balancing. A~similar study~\cite{20} was also conducted to improve the routing strategy based on OSPF for QKD networks. Likewise, based on traditional routing protocols, considering the similarity between wireless ad~hoc networks (WANETs) and QKD networks, the~authors of~\cite{21} applied three typical WANET routing protocols to QKD networks to solve the problem of frequent changes in link-state. Following a similar idea, a~dynamic routing scheme for QKD networks based on trusted relays was proposed in~\cite{22}. Some scholars have also conducted research on routing protocols of QKD networks from other aspects, such as adaption to multiple types of QKD networks~\cite{23, 24}, security enhancement~\cite{25}, and~cross-talk suppression~\cite{26}. In~summary, the~research mentioned above has not paid enough attention to high quantum key utilization, which is currently the major concern of QKD~networks.

\begin{figure}[H]
	\centering
	\includegraphics[width=10.5 cm]{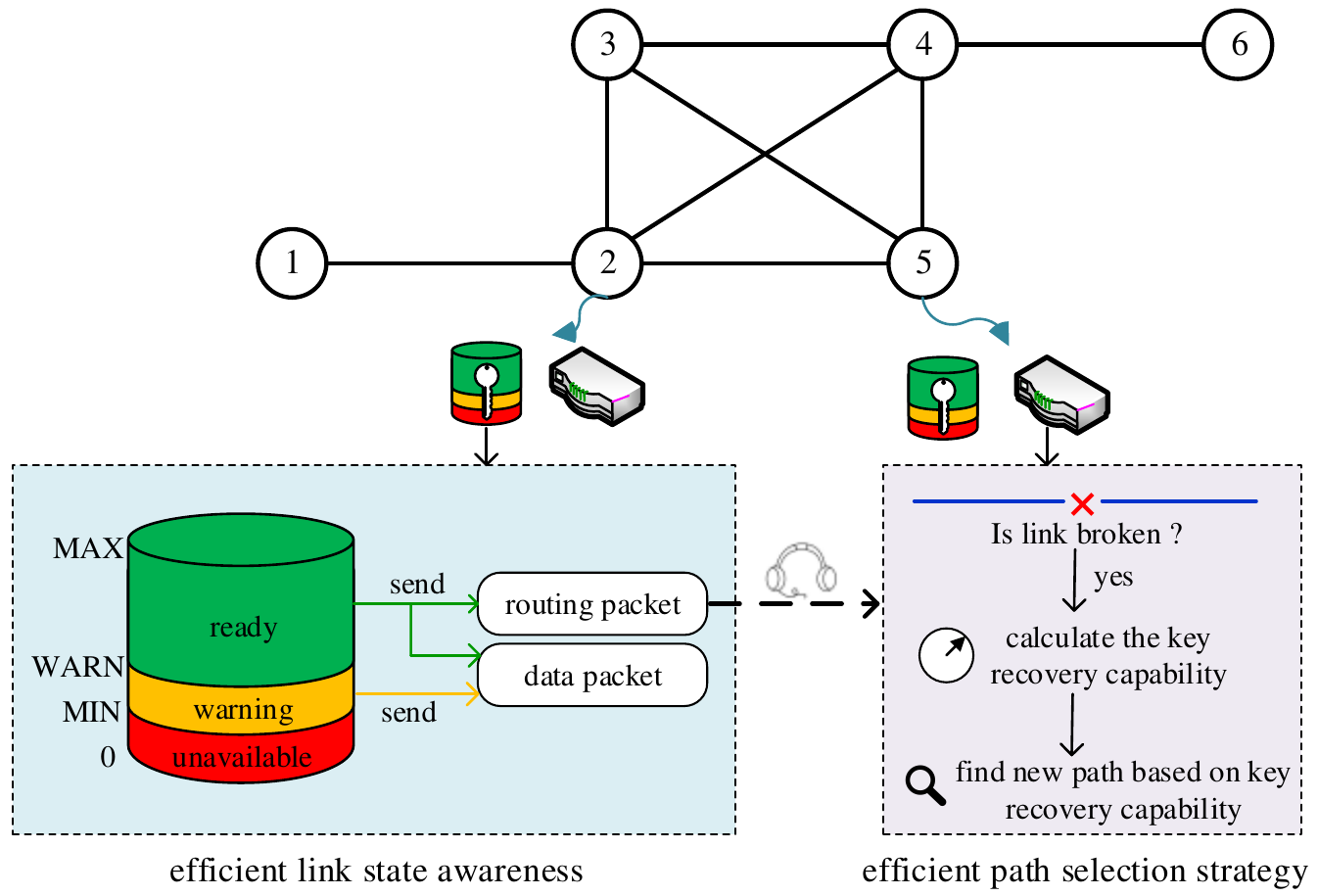}
	\caption{The architecture of~QOLSR. The numbers 1-6 represent the nodes in QKD networks.\label{fig1}}
\end{figure}

In order to achieve high quantum secure key utilization, the~problem of excessive packet loss due to frequent changes in link-state caused by the limited quantum secure key generation capability of QKD devices should be tackled first. Taking into account that both WANETs and QKD networks have frequent changes in link-state~\cite{27}, the~typical routing protocols of WANETs, including ad~hoc on-demand distance vector routing (AODV), destination sequenced distance vector (DSDV), and~optimized link-state routing (OLSR) are chosen as the research basis for designing a routing protocol for QKD networks. Both analysis and our simulation results using these three protocols show that OLSR is a more suitable option for QKD networks. To~further enhance quantum secure key utilization of QKD networks, an~efficient routing protocol based on OLSR, named QOLSR, for~QKD networks is proposed in this paper. The~graphical representation and the architecture of the proposed QOLSR protocol are shown in Figure~\ref{fig1}. First, we modify the link-state awareness mechanism in OLSR to reduce packet loss during path switching in QKD networks. A~more efficient link-state awareness mechanism is proposed in which decisions on sending data and routing packets are determined by the state of the key pool. Second, a~new routing metric based on secure key recovery capability is designed. The~suggested route selection approach based on quantum secure key recovery capability can reduce the frequency of path switching in QKD networks and increase quantum secure key utilization. Finally, simulation tests are used to evaluate the proposed QOLSR protocol. The~simulation data demonstrate the validity and efficiency of QOLSR. Moreover, the~benefit of our protocol becomes more apparent with the growth of communication~traffic.

\unskip

The remainder of this paper is organized as follows. In~the next section, Section~\ref{sec2}, the~limitations of AODV, DSDV, and~OLSR in QKD networks are first analyzed, and~then the proposed efficient routing protocol, QOLSR, is introduced in detail. Simulation results and detailed analysis is provided in Section~\ref{sec3}. Finally, the~summary of this work is drawn in Section~\ref{sec4}.

\section{The proposed QOLSR protocol}\label{sec2}
\subsection{Analysis of the limitations of AODV, DSDV, and OLSR in QKD networks}
QKD networks consume quantum secure keys on links during communication. As a~result, the~link cannot communicate normally when quantum secure key resources remaining on the link are insufficient. The~quantity of quantum secure key resources remaining on the link determines whether or not the link can continue to transfer data. Hence, the~problem of frequent changes in the link-state caused by changes in the number of remaining key resources on the link must be resolved to make QKD networks efficient. Different from wired networks but similar to WANETs, QKD networks have changes in connection state between nodes caused by node movement. Thus, inspired by~\cite{21}, three typical WANET routing protocols---AODV, DSDV, and~OLSR---are selected in this research. First, we analyze the limitations of applying the three protocols to QKD networks. Then, a~simulation experiment and performance comparison are performed using NS3 in Section~\ref{3.1}. Finally, we make further improvements based on the OLSR protocol, which has the best performance among the three~protocols.

The AODV protocol is a demand-driven passive routing protocol~\cite{28}. Routing discovery is performed by sending routing request RREQ packets and receiving routing response RREP packets when the source node has communication requirements and there is no corresponding route in the routing table. The~link-state is sensed by sending and receiving HELLO packets periodically. The~QKD network with this protocol suffers from high time delays and packet loss during the process of changing routing paths, resulting in underutilization of quantum secure keys. To~solve this problem, the~link-state should be pre-sensed by including information conveying the number of quantum secure keys in the HELLO packet. By~minimizing the time required for link-state sensing, the~performance of QKD networks can be increased. Then, the~routing metrics should be improved by adding parameters such as quantum secure key generation capability, quantum secure key consumption rate, and~the amount of remaining quantum secure key resources. Moreover, the~frequency of path switching can be reduced by increasing the working duration of the path, in~turn improving quantum secure key utilization of QKD networks. However, the~on-demand-driven passive routing discovery mode makes it difficult to optimize the time necessary for routing~discovery.

The DSDV protocol~\cite{29} is an active routing protocol that is driven by routing tables. The~routing table is broadcast on a regular basis to detect changes in global topology. Each node in the network sends its routing table to other nodes during the broadcasting process, resulting in a large routing cost. Since the routing packet also needs to be protected by QKD encryption, a~huge number of quantum secure keys will be consumed. In~large-scale QKD networks, where quantum secure keys are valuable, this cannot be tolerated. When the current communication path is disabled, substantial packet loss occurs in QKD networks using the DSDV protocol since the loss cannot be detected until the next broadcast cycle. Therefore, DSDV should be optimized by shortening the broadcast cycle and reducing the routing overhead. However, these goals are mutually constraining and difficult to optimize~simultaneously. 

Likewise, the~OLSR protocol is an active routing protocol driven by routing tables~\cite{30}. OLSR combines global topology awareness based on periodic broadcast as in DSDV and link-state awareness based on HELLO packet as in AODV through multipoint replay (MPR). It not only ensures timely awareness of global topology but also significantly reduces routing overhead. However, QKD networks with OLSR protocol also suffer from excessive time delays and packet loss caused by the link-state awareness mechanism based on HELLO packets and the routing metrics based on minima-hopping during path switching. Therefore, OLSR can be optimized by adding the information of the remaining quantum secure keys in the HELLO packet to pre-sense the link state. In~addition, parameters such as quantum secure key generation capabilities should be added to the routing metrics. Unlike AODV, the~routing table of OLSR is constructed locally using topology awareness and does not require the routing discovery~procedure.

Based on the above analysis, OLSR is the most suitable of the three protocols for QKD networks. This is also confirmed by the simulation results in Section~\ref{3.1}. Therefore, a~more efficient routing protocol based on OLSR for QKD networks is investigated in this~research.

\subsection{An overview of the OLSR routing protocol}
\begin{figure}[H]
	\centering
	\includegraphics[width=10.5 cm]{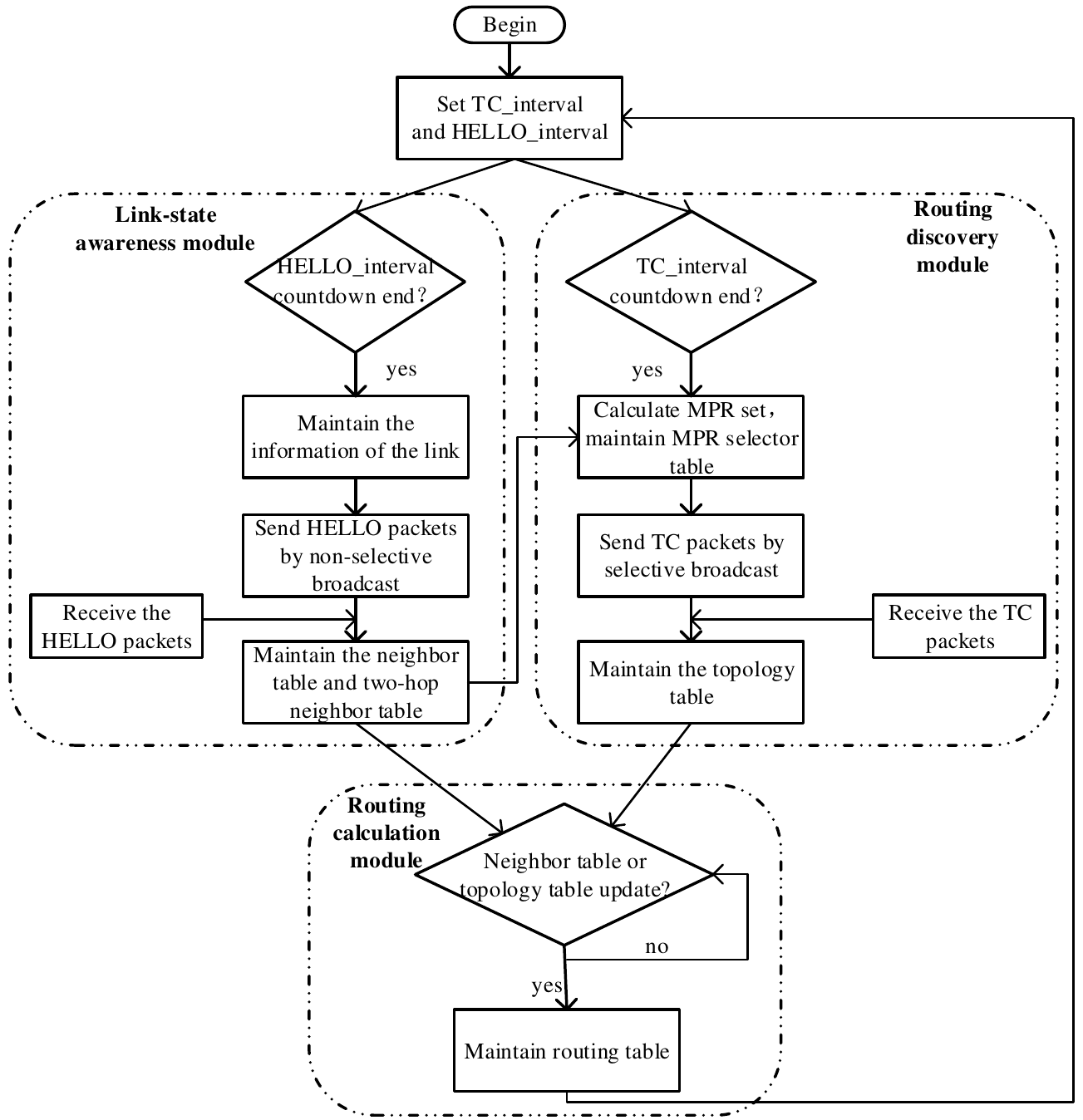}
	\caption{Schematic diagram of OLSR routing protocol\label{fig2}}
\end{figure}

Figure~\ref{fig2} depicts the OLSR protocol’s workflow. As~shown, it includes three main modules: link-state awareness, routing discovery and routing calculation modules~\cite{30}. HELLO packets and TC packets are used for awareness of the link-state and the topology, respectively. HELLO\_interval and TC\_interval represent the sending periods of HELLO and TC packets, respectively. The~shorter the sending periods, the~higher the sensitivity. As~a result, there is a lot of routing overhead and the over-consumption of quantum secure keys. Each module’s workflow will be discussed in depth below. The~limits of OLSR in QKD networks, as~well as optimization strategies, are also~discussed.

The link-state awareness mechanism of OLSR is similar to that of AODV. Changes in link-state are sensed by sending and receiving HELLO packets periodically. In~order to calculate the MPR set, the~HELLO packages in OLSR must not only sense the link-state among neighbor nodes but also undertake the task of calculating the two-hop neighbors. Therefore, HELLO packets containing the information of all neighbor nodes are sent to each neighbor node by non-selective broadcast. Each node in the network maintains the neighbor table and two-hop neighbor table by listening to the HELLO packets. Because~there is no pre-sense of link-state in OLSR, path-switching causes significant time delays and packet loss. The~link-state can be pre-sensed by adding information about the remaining quantum secure keys in the HELLO packets. Thereby, the~utilization of quantum secure keys can be enhanced by reducing the packet loss caused by path~switching.

\unskip

The routing discovery module calculates the MPR set in real-time according to the neighbor table and the two-hop neighbor table. The~MPR mechanism significantly reduces the routing overhead and has no effect on the awareness of global topology. Since the topology changes dynamically, each node in the network needs to store an MPR selector table that contains the information of the real-time MPR set. Subsequently, the~TC packets containing the MPR selector table will be sent to each neighbor node by selective broadcast based on the MPR mechanism. In~the meantime, each node in the network maintains the topology table by listening to the TC packets. The~topology table is used to find the source node based on the destination node and then calculate the optimal~path.

The routing calculation module generates a routing table based on the neighbor table and topology table generated by the above two modules. The~OLSR routing protocol will recalculate the routing table when the neighbor table or topology table is changed. OLSR selects the optimal path based on the minima-hopping method. However, information transmission in QKD networks relies heavily on quantum secure keys. Therefore, the~path’s quality is determined by the information of the quantum secure key. The~routing metrics should be improved by adding parameters such as quantum secure key generation capability, quantum secure key consumption rate, and~the amount of remaining quantum secure key resources in order to improve quantum secure key utilization of QKD~networks.

\subsection{The proposed QOLSR protocol}
According to the previous analysis, OLSR is a better fit for QKD networks than AODV or DSDV. Therefore, an~optimized routing protocol based on OLSR that is more efficient and more suitable for QKD networks is proposed in this study. The~main drawbacks of QKD networks are the frequent changes in link-state, which cause low quantum secure key utilization. The~proposed routing protocol, QOLSR, can reduce the frequency of changes in link-state and enhance quantum secure key utilization of QKD networks through link-state awareness and a path optimization~algorithm.

\subsubsection{Efficient link-state awareness mechanism for QKD network}
The traffic on each link is dynamic due to the high concurrency and highly fluctuating nature of communication requirements. Furthermore, each link’s key generation capability varies, resulting in differences in the quantity of residual key resources on each link throughout communication. The~link-state in QKD networks depends on whether there are enough quantum secure key resources in the key pool for communication. During~communication, the~variable amount of remaining quantum secure key resources regularly causes changes in the link-state. In~turn, the~topology of QKD networks changes. Link-state awareness must be timely and efficient to overcome the challenge of frequent topology changes. By~utilizing a timely and effective link-state awareness mechanism, QKD networks can communicate with high quantum secure key~utilization.

As mentioned above, in~order to enhance quantum secure key utilization in QKD networks, a~key pool is set on each link. Considering the limited storage resources, the~amount of quantum secure key resources in each key pool has a maximum threshold MAX. A~minimal threshold MIN is specified for each key pool to ensure that there are enough authentication keys for key generation. In~addition, each key pool also has a warning threshold WARN for link-state awareness. WARN can warn that the link is going to be in a broken state and a new routing path should be calculated in advance. According to the three thresholds, there are three states of the key pool: unavailable, warning, and~ready, as~shown in Figure~\ref{fig1}. In~this paper, the~link-state is judged through the states of the key pool. The~ready state of the key pool indicates that the link is in a healthy communication state. The~warning state of the key pool indicates that the link is about to transform to a broken state. The~unavailable state of the key pool indicates that the link is in a broken~state.

\begin{figure}[H]
	\centering
	\includegraphics[width=10.5 cm]{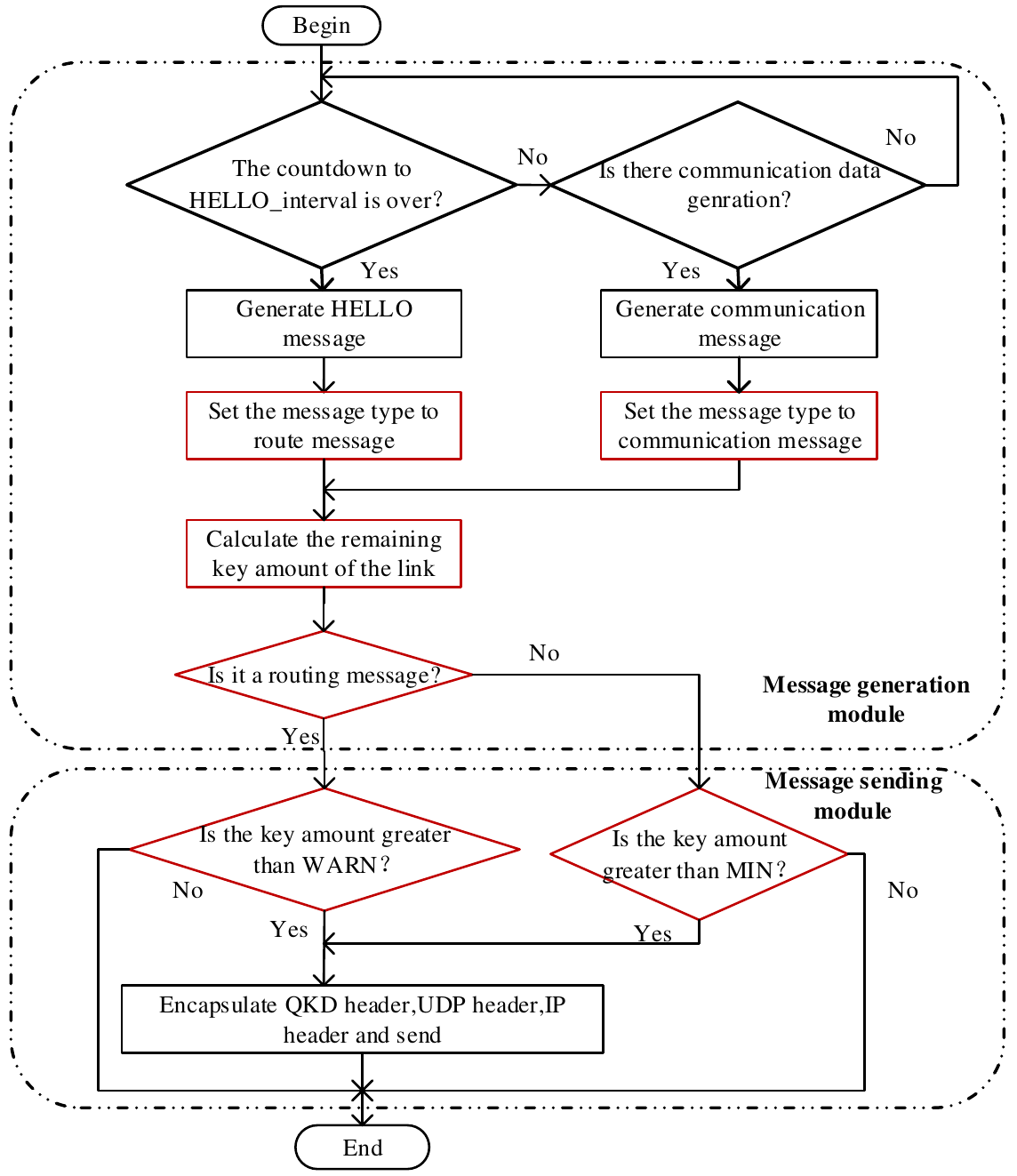}
	\caption{The link-state prediction scheme\label{fig3}}
\end{figure}

In the OLSR protocol, a~new routing path is calculated locally when the neighbor routing table or the topology table changes. When the link is disconnected, the~neighbor routing table will be changed due to the loss of HELLO packets. Subsequently, the~two-hop neighbor table changes, too. This causes further changes in the MPR selector table and topology table. Therefore, timely awareness of the link-state is largely dependent on when HELLO packet loss is sensed. If~the HELLO packets sent by neighbor nodes are not received, the~OLSR protocol assumes the link is broken due to inadequate quantum secure keys. To~communicate the information of broken links to each node in the entire network, numerous TC interval cycles are required. Data packets cannot be transferred normally during this time because the routing path in the routing table is actually inaccessible. To~address this issue, this paper proposes a pre-sensed link-state strategy based on the amount of remaining quantum secure key resources, which involves separating communication data packets from routing packets and upgrading the QKD network’s HELLO packet transmission mechanism. The~workflow of this scheme is shown in Figure~\ref{fig3}. In~comparison to the original link-state awareness mechanism of OLSR, the~red component is the improved version proposed in this~paper.

The required key pool's information can be retrieved locally using HELLO packets. Therefore, whether the number of remaining quantum secure key resources can meet communication needs can be estimated locally. WARN can anticipate the link-state by comparing the quantity of remaining quantum secure key resources with the warning threshold. If~the quantity of remaining quantum secure key resources is less than WARN, the~node will not broadcast HELLO packets to neighboring nodes. Communication data packets are sent normally until the amount of remaining quantum secure key resources is lower than the minimum threshold MIN or the communication path is changed. Based on the above scheme, there will be WARN-MIN quantum secure keys that can be used to provide communication services during path switching, which can enhance quantum secure key utilization in QKD~networks.

\unskip

\subsubsection{Efficient optimization path selection strategy for QKD network}
In addition to efficient link-state awareness as proposed above, QKD networks should also be able to make timely path adjustments based on this awareness. OLSR selects the optimal path based on the minima-hopping method. However, QKD networks are different from traditional networks. Information transmission in QKD networks relies heavily on quantum secure keys. As~a result, the~path’s quality is determined by the remaining quantum secure key resources, the~ability to generate quantum secure keys, and~the quantum secure key consumption rate. This research assesses the optimal path in terms of the remaining quantum secure key resources, the~ability to generate quantum secure keys, and~the quantum secure key consumption rate, and~proposes an efficient path selection strategy based on key recovery~ability.

In order to reduce the switching frequency of the communication path, the~path with the longest sustainable working time is considered the optimal path. The~sustainable working time takes into account the amount of remaining quantum secure key resources, quantum secure key generation capability, and~quantum secure key consumption rate. Given the topology of QKD networks, $N = \left( {V,E} \right)$, where $V$ is the set of nodes and $E$ is the set of edges. The~quantum secure key rate on each edge is represented by $\left\{ {{C_{\left( {u,v} \right)}}\mid \left( {u,v} \right) \in E} \right\}$, where $\kappa$ is packet size, $\omega _{\left( {u,v} \right)}^{s,t}\left( i \right)$ is traffic on edge $\left( {u,v} \right)$ at time $i$ when source node $s$ communicates with destination node $t$, and~$Cu{r_{\left( {u,v} \right)}}\left( i \right)$ and $Du{r_{\left( {u,v} \right)}}\left( i \right)$ represent the amount of remaining available quantum secure key resources and the sustainable working time of edge $\left( {u,v} \right)$ at time $i$, respectively. When the key generation rate is higher than the key consumption rate, the~sustainable working time of the link is infinite. When the key generation rate is lower than the key consumption rate, the~formula to calculate $Du{r_{\left( {u,v} \right)}}\left( i \right)$ is:

\begin{equation}
Du{r_{\left( {u,v} \right)}}\left( i \right) = \frac{{Cu{r_{\left( {u,v} \right)}}\left( i \right)}}{{\sum\limits_{s \in V',t \in V'} {\left( {\omega _{\left( {u,v} \right)}^{s,t}\left( i \right) \cdot \kappa } \right)}  - {{C'}_{\left( {u,v} \right)}}}},
\end{equation}

where $\sum\limits_{s \in V',t \in V'} {\left( {\omega _{\left( {u,v} \right)}^{s,t}\left( i \right) \cdot \kappa } \right)} $ is used to calculate the total traffic of each communication pair through edge $\left( {u,v} \right)$ at time $i$. The~sustainable working time is calculated based on the current routing protocol. Therefore, if~the routing protocol is adjusted according to the calculation result, it will cause uncontrollable negative feedback problems, resulting in unstable communication and low quantum secure key~utilization.

Thus, this paper suggests using quantum secure key recovery time instead of sustainable working time as a new routing metric. Specifically, sustainable working time refers to the time required for the quantum secure keys in the key pool to be exhausted when the quantum secure key consumption rate is higher than the quantum secure key generation rate. Key recovery time refers to the time required for the key pool to be filled when the quantum secure key generation rate is higher than the quantum secure key consumption rate. With~a maximum threshold in the key pool, the~amount of quantum secure key resources cannot increase indefinitely. Therefore, the~shorter the key recovery time, the~closer the key pool is to the full state, and~the higher the priority of the link. On~the contrary, the~longer the quantum secure key recovery time, the~closer the key pool is to the empty state, and~the lower the priority of the~link.

Considering that the link priority is inversely proportional to the quantum secure key recovery time, the~reciprocal of the quantum secure key recovery time (called the quantum secure key recovery capability) is used to judge link priority. Let $Ma{x_{\left( {u,v} \right)}}$ represent the maximum threshold of the key pool on edge $\left( {u,v} \right)$ and ${\gamma _{\left( {u,v} \right)}}\left( i \right)$ represent the quantum secure key recovery capability of edge $\left( {u,v} \right)$ at time $i$. The~formula to calculate ${\gamma _{\left( {u,v} \right)}}\left( i \right)$ is:

\begin{equation}
{\gamma _{\left( {u,v} \right)}}\left( i \right) = \frac{{{C_{\left( {u,v} \right)}} - \sum\limits_{s \in V',t \in V'} {\left( {\omega _{\left( {u,v} \right)}^{s,t}\left( i \right) \cdot \kappa } \right)} }}{{Ma{x_{\left( {u,v} \right)}} - Cu{r_{\left( {u,v} \right)}}\left( i \right)}},
\end{equation}

In addition, for~a path that contains multiple links in the QKD network, the~priority of the path can be expressed by the minimum value of the priority of all the links it contains. The~ formula is:

\begin{equation}
{\gamma _{path}}\left( i \right) = \mathop {\min }\limits_{\left( {u,v} \right) \in path} \frac{{{C_{\left( {u,v} \right)}} - \sum\limits_{s \in V',t \in V'} {\left( {\omega _{\left( {u,v} \right)}^{s,t}\left( i \right) \cdot \kappa } \right)} }}{{Ma{x_{\left( {u,v} \right)}} - Cu{r_{\left( {u,v} \right)}}\left( i \right)}},
\end{equation}

Path optimization based on the quantum secure key recovery capability has a higher level of stability and is also good at coping with the link-state frequency changes in QKD networks. Therefore, quantum secure key utilization can be enhanced by selecting the optimal~path.

\subsubsection{An Example of QOLSR~Protocol}\label{2.3.3}
\begin{figure}[H]
	\centering
	\includegraphics[width=10.5 cm]{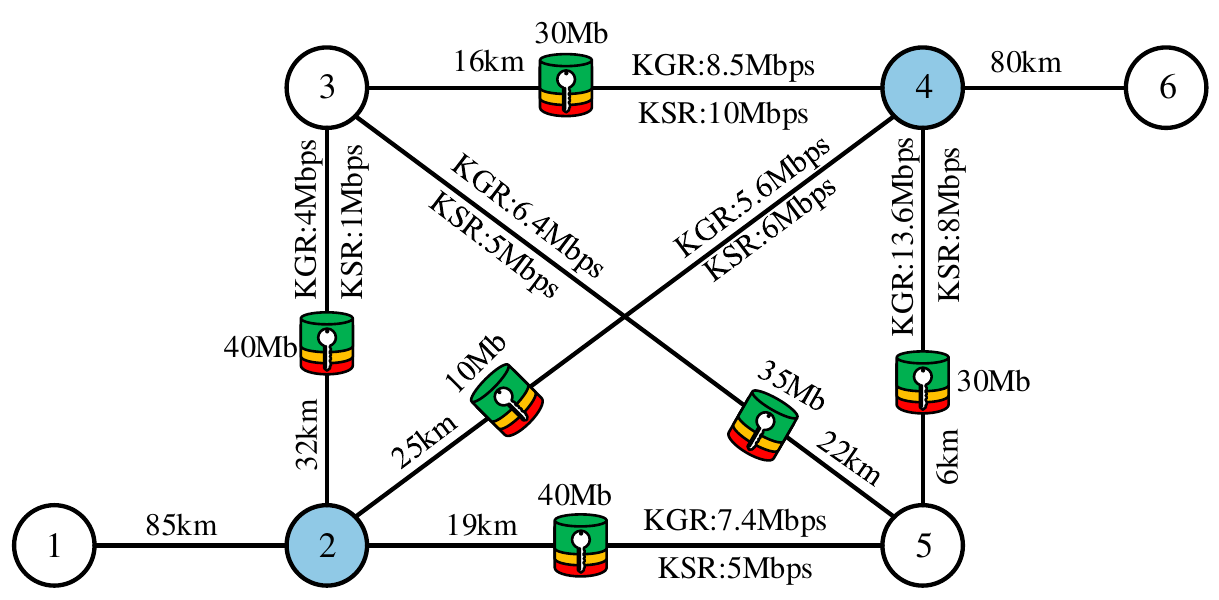}
	\caption{An  example of QOLSR: KGR, quantum secure key generation rate; KSR, quantum secure key consumption~rate. The numbers 1-6 represent the nodes in QKD networks.\label{fig4}}
\end{figure}

\begin{figure}[H]
	\centering
	\includegraphics[width=10.5 cm]{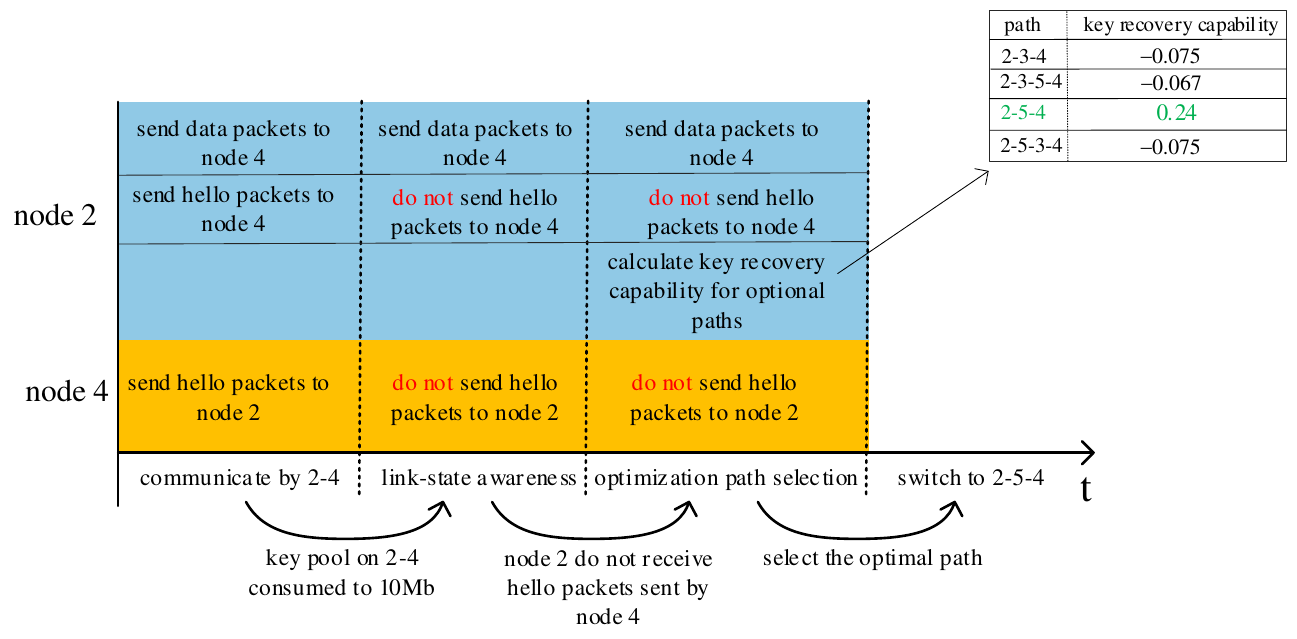}
	\caption{Workflow of QOLSR in this~example.\label{fig5}}
\end{figure}

This section provides a basic example to explain QOLSR protocol's link-state awareness mechanism and path optimization. In~this section, an~SECOQC network is chosen as the network topology, as~illustrated in Figure~\ref{fig4}. The~length of the optical fiber link is indicated by the numbers on each edge (in km). It is assumed that each edge generates the quantum secure key through the VWDDS protocol. The~optical parameters of the QKD devices are the same as in~\cite{32}. The~GLLP formula can be used to calculate the key generation rate on each link, as~shown in Figure~\ref{fig4}. The~key pool parameter MIN is set to 2~Mbit, WARN to 10~Mbit, and~MAX to 50~Mbit.
It is assumed that node 2 and node 4 communicate by path 2{-}4 at the beginning. Link 2{-}4 has a key generation rate of 5.6 Mbps and a key consumption rate of 6 Mbps. Figure~\ref{fig5} shows the entire workflow of link-state awareness and path optimization of QOLSR. First, since the key consumption rate of link 2{-}4 is higher than the key generation rate, the~keys in key pool 2{-}4 will be consumed indefinitely. When the number of keys in key pool 2{-}4 decreases to the warning threshold of 10~Mb, the~link-state awareness mechanism of QOLSR will be triggered. At~the same time, HELLO packets will not be sent between nodes 2 and 4. However, the~data packets will be sent as usual. When node 2 does not receive HELLO packets from node 4 during the HELLO interval, link 2{-}4 is considered disconnected. As~a result, it will begin routing discovery in order to find a new communication path. Packet loss during path switching is reduced and quantum secure key consumption is improved because data packets are still transferred properly while searching for a new~path.

\unskip

\unskip

There are four possible communication paths between nodes 2 and 4, as~shown in Figure~\ref{fig4}: 2{-}3{-}4, 2{-}3{-}5{-}4, 2{-}5{-}4, and~2{-}5{-}3{-}4. Thus, the~optimal path will be selected based on key recovery capability. First, we calculate the key recovery capability of path 2{-}3{-}4. Path 2{-}3{-}4 consists of two links: 2{-}3 and 3{-}4. Therefore, the~key recovery capabilities of links 2{-}3 and 3{-}4 are calculated first, and~then the key recovery capability of path 2{-}3{-}4 is calculated as the minimum value between them. The~key generation rate on link 2{-}3 is 4~Mbps, the~key consumption rate is 1~Mbps, and~the remaining key amount in the key pool is 40~Mb, as~shown in Figure~\ref{fig4}. Therefore, its key recovery capability is $\frac{{4~\text{Mbps} - 1~\text{Mbps}}}{{50~\text{Mb} - 40~\text{Mb}}} = 0.3$. Similarly, the~key recovery capability of link 3-4 is $\frac{{8.5~\text{Mbps} - 10~\text{Mbps}}}{{50~\text{Mb} - 30~\text{Mb}}} =  - 0.075$. Therefore, the~key recovery capability of path 2{-}3{-}4 is $-$0.075. Similarly, the~key recovery capabilities of paths 2{-}3{-}5{-}4, 2{-}5{-}4, and~2{-}5{-}3{-}4 can be calculated as shown in Figure~\ref{fig5}. As~a result, it can be seen that path 2{-}5{-}4 has the best key recovery capability; hence it is chosen as the new communication path. Since the key recovery capability of path 2{-}5{-}4 is higher, it can continue to function properly for a longer period of time. Subsequently, the~frequency of path switching is lowered, and~quantum secure key utilization~improves.

\section{Simulation results and analysis}\label{sec3}
\subsection{Comparative study of AODV, DSDV and OLSR}\label{3.1}
\begin{figure}[H]
	\centering
	\includegraphics[width=10.5 cm]{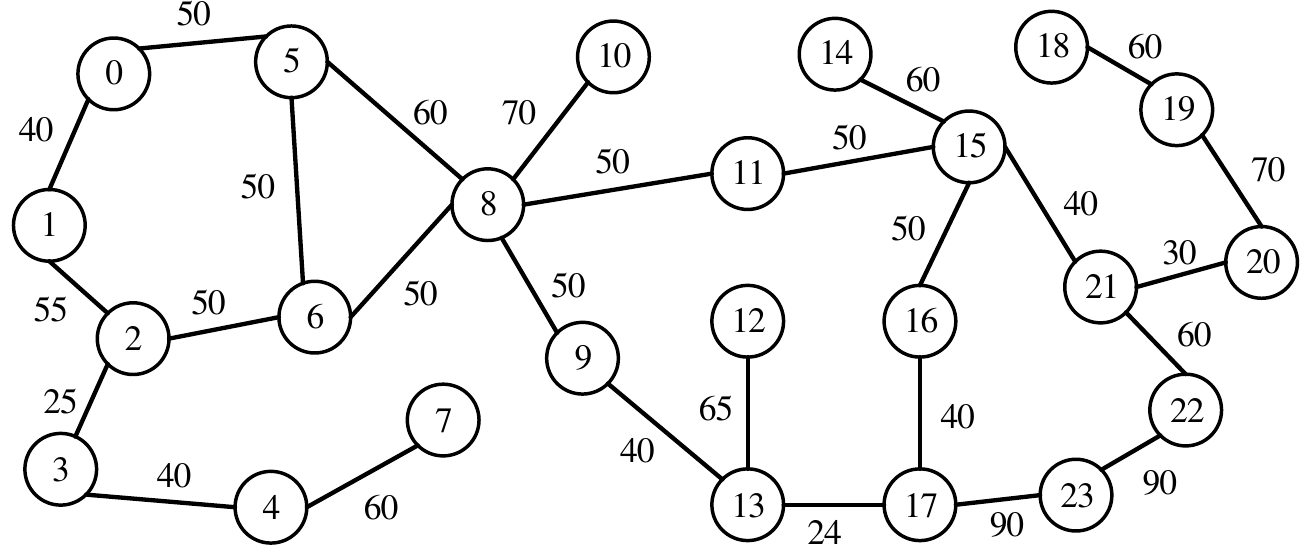}
	\caption{Topology of the USNET network~\cite{31}. The numbers on each edge represent the length of the optical fiber link (in km).\label{fig6}}
\end{figure}

A comparative study is done with AODV, DSDV, and~OLSR through simulator NS3. In~this paper, the~QKD simulation module based on NS3 developed in the literature~\cite{21} is used to simulate the QKD network. The~open source code with this QKD simulation module can be downloaded from the git repository supplied in~\cite{21}. A~modified USNET network is selected as the network topology in this experiment, as~shown in Figure~\ref{fig6}. There are 24~nodes and 27 edges in this network topology~\cite{31}. The~numbers on each edge represent the length of the optical fiber link (in km). The~parameters of QKD devices and key pools are the same as in Section~\ref{2.3.3}. We first discuss the concept of communication level before moving on to the network requirement configuration in this experiment. The~communication level is defined as follows:

\begin{equation}
	{\rm{communication\_level = }}\frac{{{\rm{communication \_demand}}}}{{ITS\_communication\_capability}}.
\end{equation}

The highest communication demand that the network topology can theoretically accommodate is its $ITS\_communication\_capacity$~\cite{32}. If~the network’s communication demand exceeds the $ITS\_communication\_capacity$, 
severe network congestion will occur. Please refer to~\cite{32} for a detailed calculation of the $ITS\_communication\_capacity$. The~communication level is calculated as the ratio of the average communication requirement and the $ITS\_communication\_capacity$. The~QKD network cannot work normally when the average communication requirement exceeds the $ITS\_communication\_capacity$. Therefore, the~range of values of the ITS communication level is [0, 1]. In~this experiment, the~entire network communication mode is used for communication settings in the simulation process. In~other words, there is a communication requirement between any two nodes in the network. The~communication level in this experiment is set to 0.01, 0.2, 0.4, 0.6, 0.8, and~1. The~packet size is set to 500 bytes, and~the simulation time is 100 s. Due to path switching, some packets will be lost throughout the network’s communication process. Since the lost data packets are also encrypted by a quantum secure key, the~quantum secure key utilized by the lost data packets appears to have been ineffectively employed. As~a result, quantum secure key utilization refers to the fraction of quantum secure keys consumed by successfully transmitted data packets compared to the total number of quantum secure keys consumed, calculated as follows:

\begin{equation}
	QKU = \frac{{DPS}}{{TPS}},
\end{equation}

where $DPS$ and $TPS$ represent the number of quantum secure keys consumed by delivered data packets and total data packets, respectively, and $QKU$ stands for quantum secure key utilization. Quantum secure key utilization is mainly affected by the packet delivery rate (PDR), as~can be observed from the formula. Therefore, real-time PDR in the 100~s simulation is also used in this experiment to reflect the evolution of quantum secure key~utilization.

\unskip

\begin{figure}[H]
	\centering
	\subfloat[\centering]{
		\label{fig7-1}
		\includegraphics[scale=0.04]{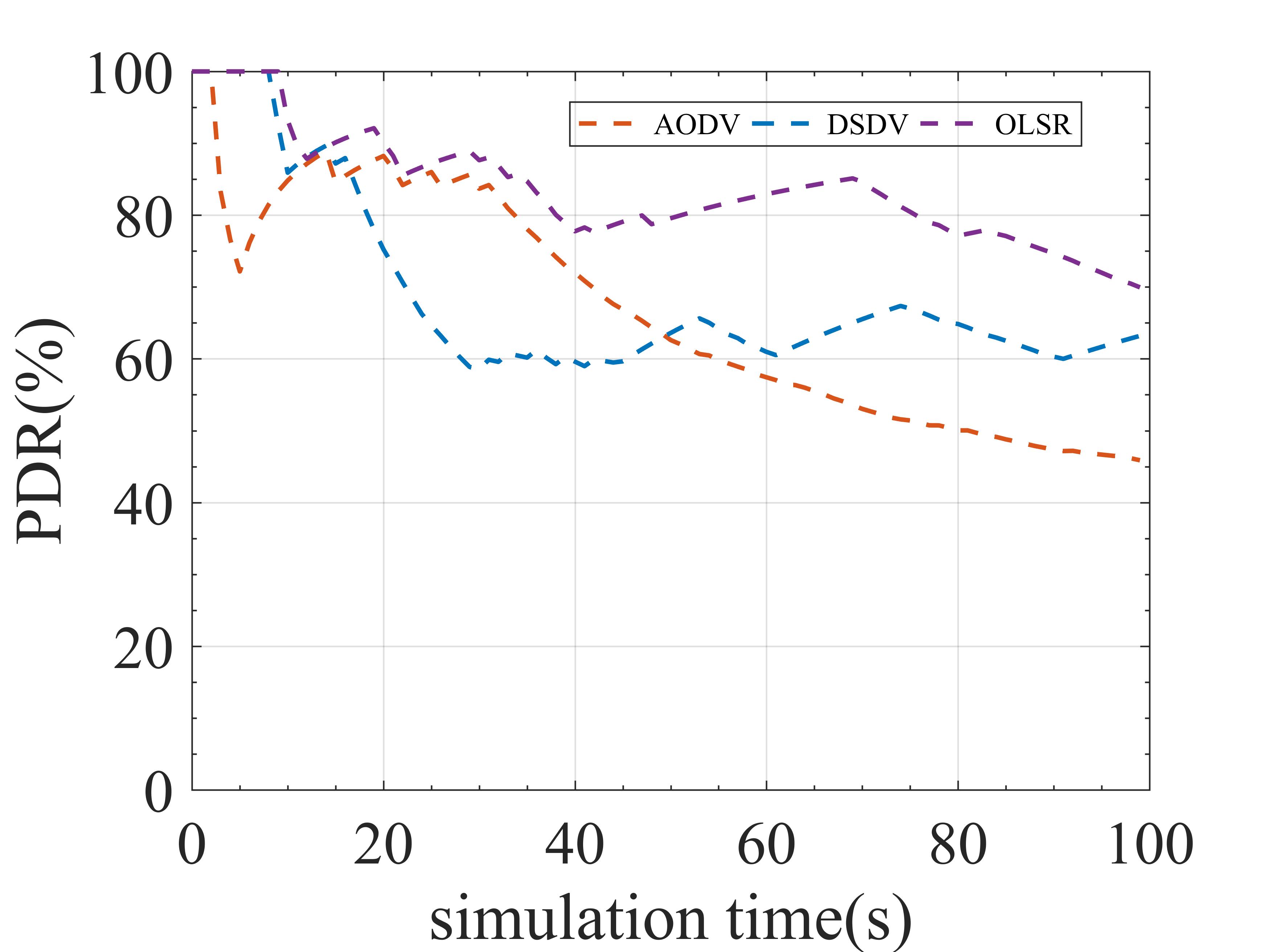}}
	\subfloat[\centering]{
		\label{fig7-2}
		\includegraphics[scale=0.04]{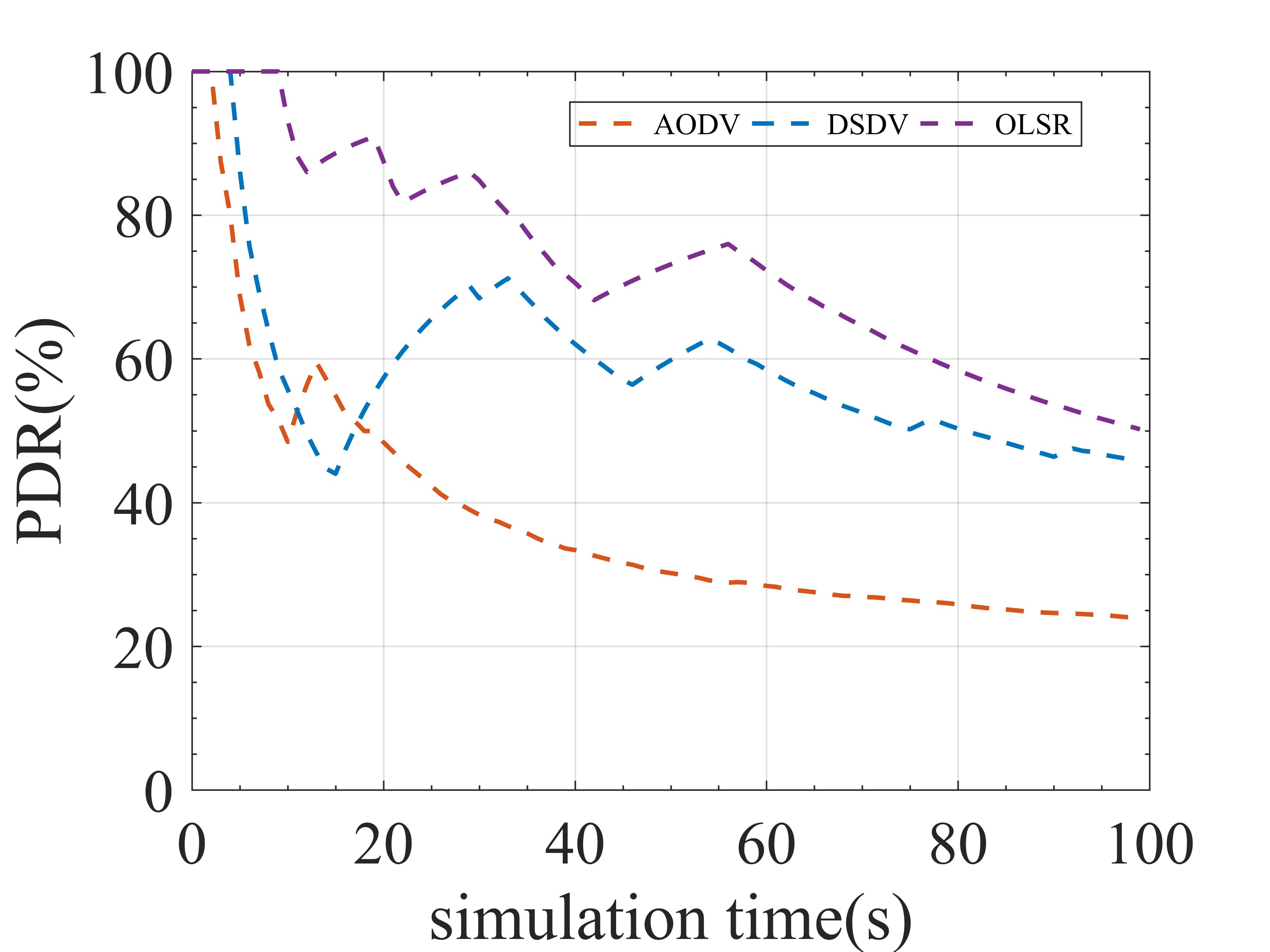}}\\ 
	\subfloat[\centering]{
		\label{fig7-3}
		\includegraphics[scale=0.04]{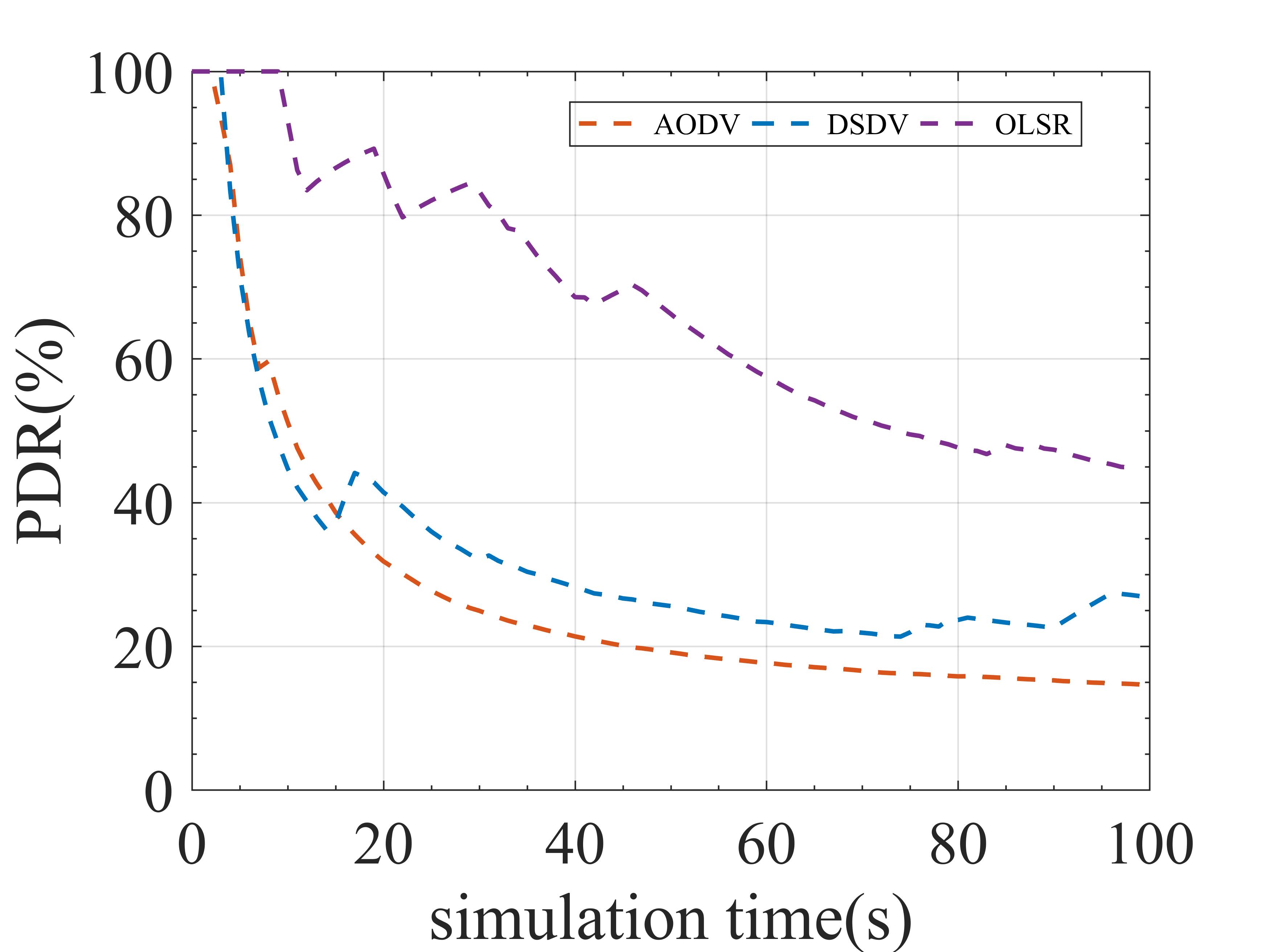}}
	\subfloat[\centering]{
		\label{fig7-4}
		\includegraphics[scale=0.04]{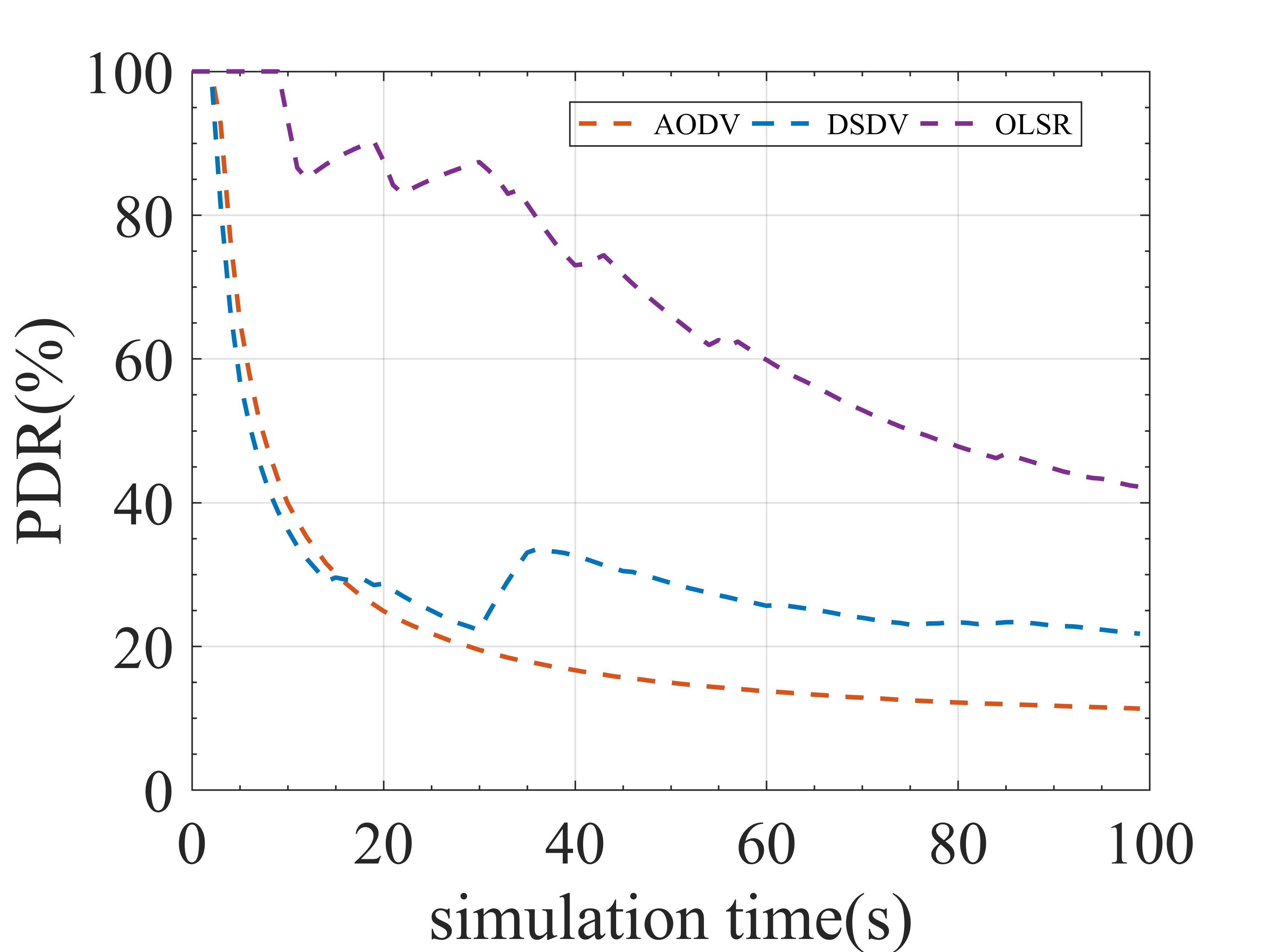}}
	\caption{Comparison of AODV, DSDV, and~OLSR in real-time~PDR. (\textbf{a}) Communication level = 0.2. (\textbf{b}) Communication level = 0.4. (\textbf{c}) Communication level = 0.6. (\textbf{d}) Communication level = 0.8.}
	\label{fig7}
\end{figure}

\begin{figure}[H]
	\centering
	\includegraphics[width=10.5 cm]{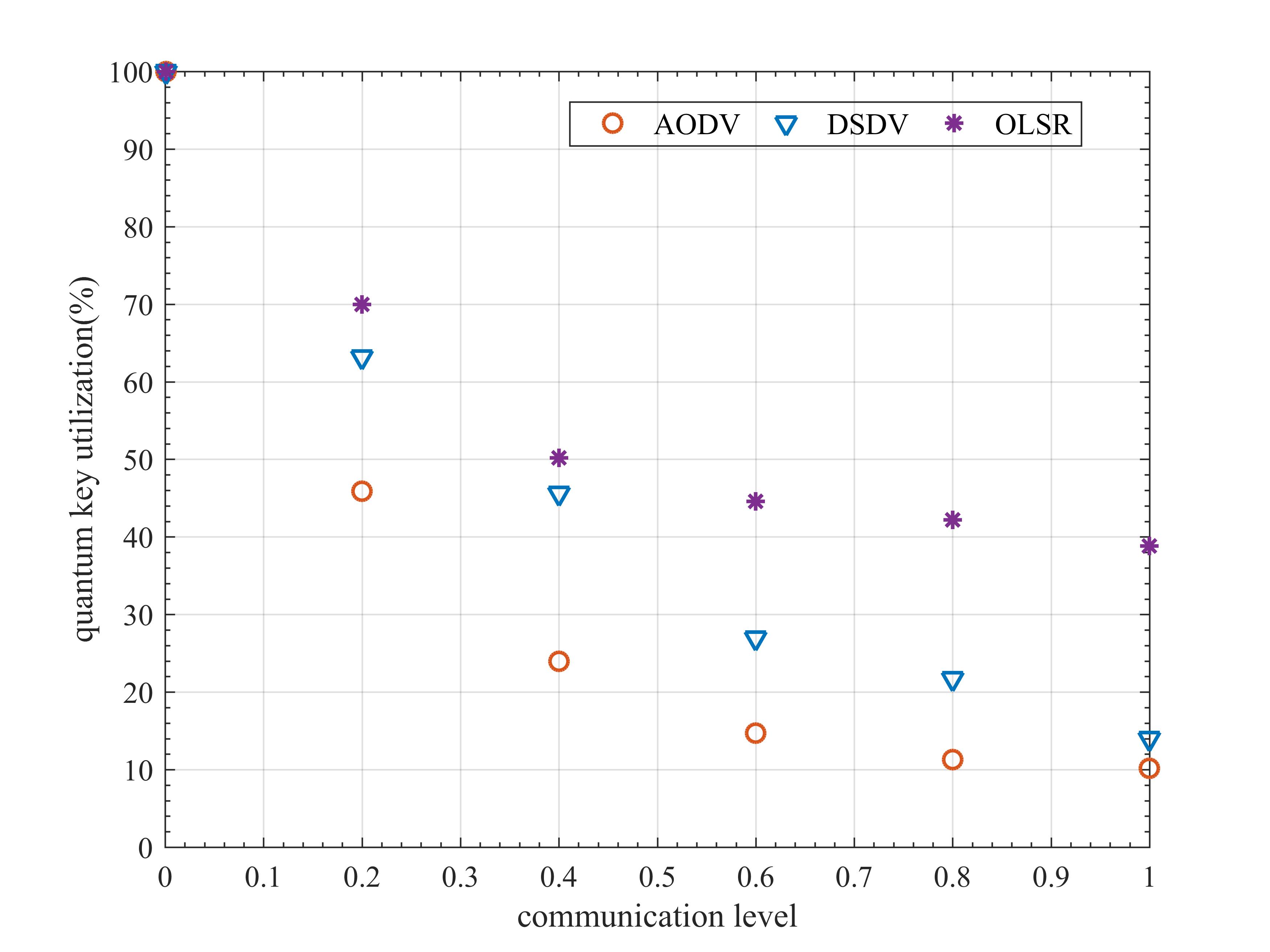}
	\caption{Comparison of AODV, DSDV, and~OLSR in quantum secure key~utilization.\label{fig8}}
\end{figure}

Figure~\ref{fig7} shows that AODV performs the worst among the three protocols, whereas the OLSR protocol performs the best. The~DSDV protocol performs better when the communication level is low. However, as~the communication level rises, its PDR drops sharply. The~variation of the PDR curve also shows that the OLSR protocol recovers the PDR faster than AODV and DSDV, implying that OLSR has better link-state awareness. The~statistical experimental results shown in Figure~\ref{fig8} indicate that quantum secure key utilization of OLSR is significantly higher than that of AODV and DSDV at all communication levels. The~performance of the OLSR protocol is mainly attributed to its dynamic link state awareness and selective broadcasting. It can detect changes in network topology faster than AODV and DSDV protocols, thereby reducing the loss of data packets during path switching. As~a result, a~higher PDR is obtained. At~the same time, fewer quantum secure keys are wasted, resulting in increased quantum secure key utilization. Results of the experiment show that OLSR is more suitable for QKD networks than the other two protocols. However, the~quantum secure key utilization of OLSR is only 38.81$\%$ when the communication level is 1, indicating that there remains significant room for improvement. As~a result, in~this paper, we attempt to propose a more efficient routing protocol based on~OLSR.

\unskip

\unskip

\subsection{Simulation results of QOLSR and analysis}\label{3.2}
In this section, we compare our proposed approach with original OLSR and the multi-SPF routing protocol proposed in~\cite{22}. The~simulation environment in Section~\ref{3.2} is the same as in Section~\ref{3.1}. 

\begin{figure}[H]
	\centering
	\subfloat[\centering]{
		\label{fig9-1}
		\includegraphics[scale=0.04]{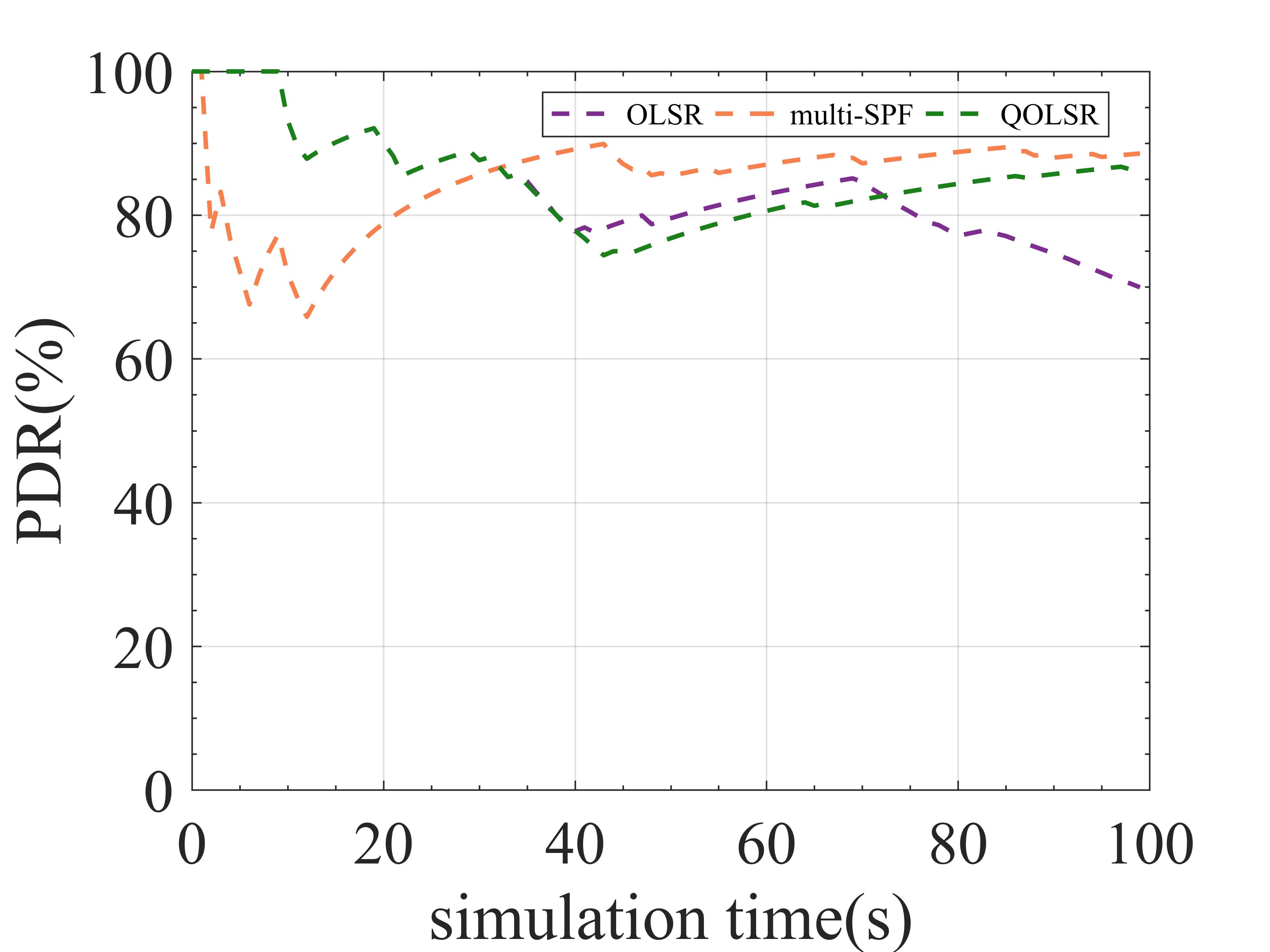}}
	\subfloat[\centering]{
		\label{fig9-2}
		\includegraphics[scale=0.04]{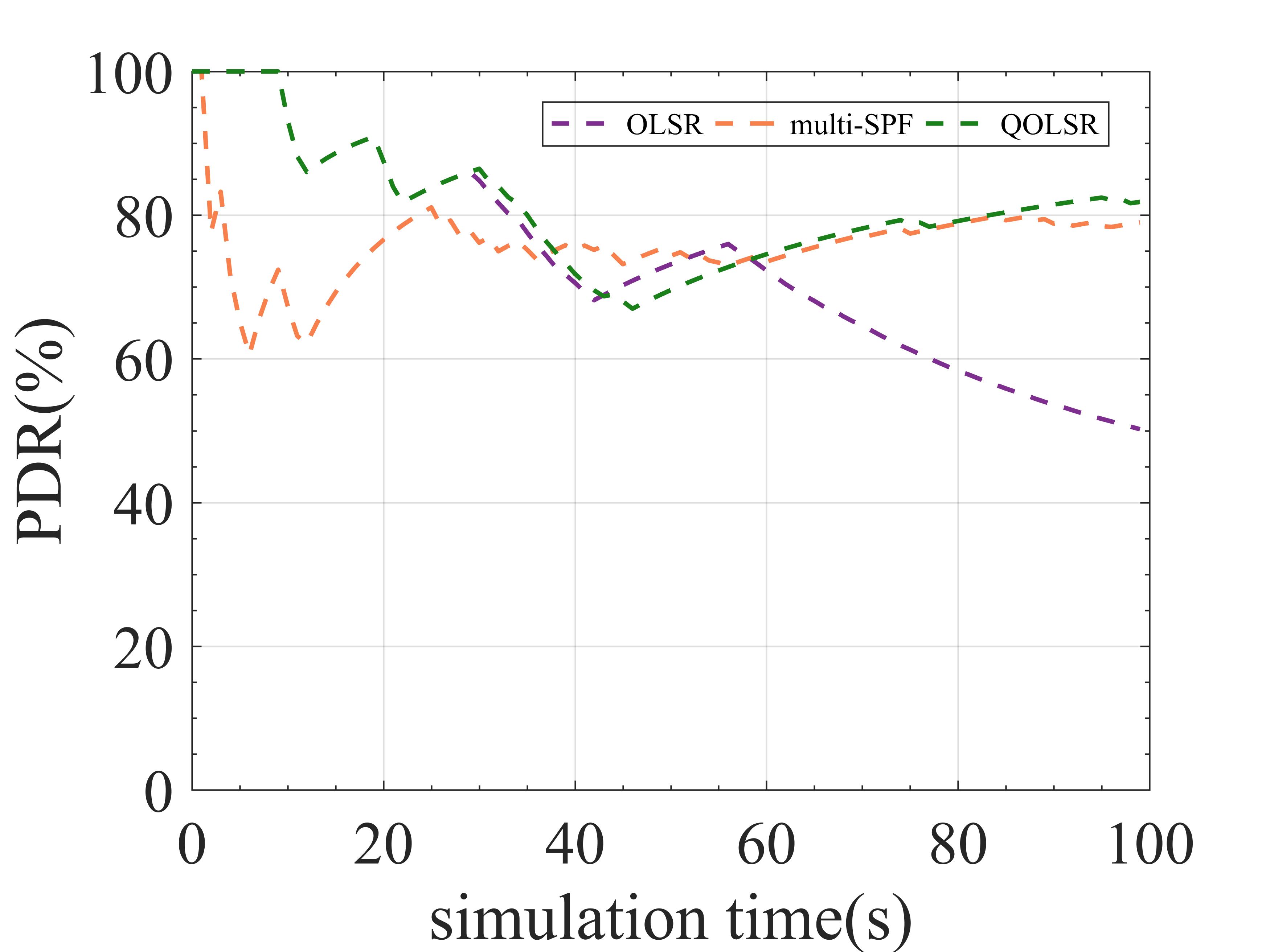}}\\ 
	\subfloat[\centering]{
		\label{fig9-3}
		\includegraphics[scale=0.04]{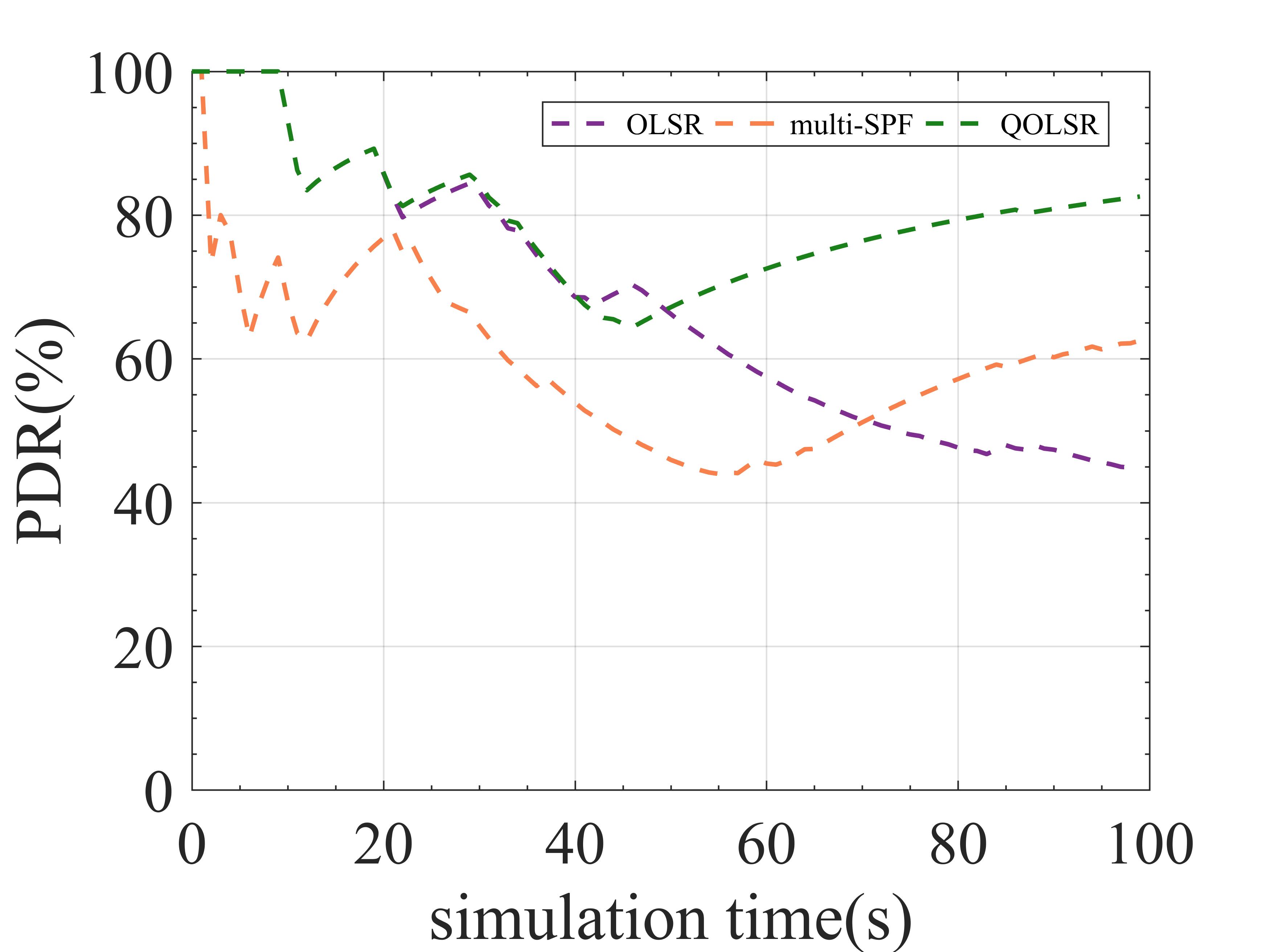}}
	\subfloat[\centering]{
		\label{fig9-4}
		\includegraphics[scale=0.04]{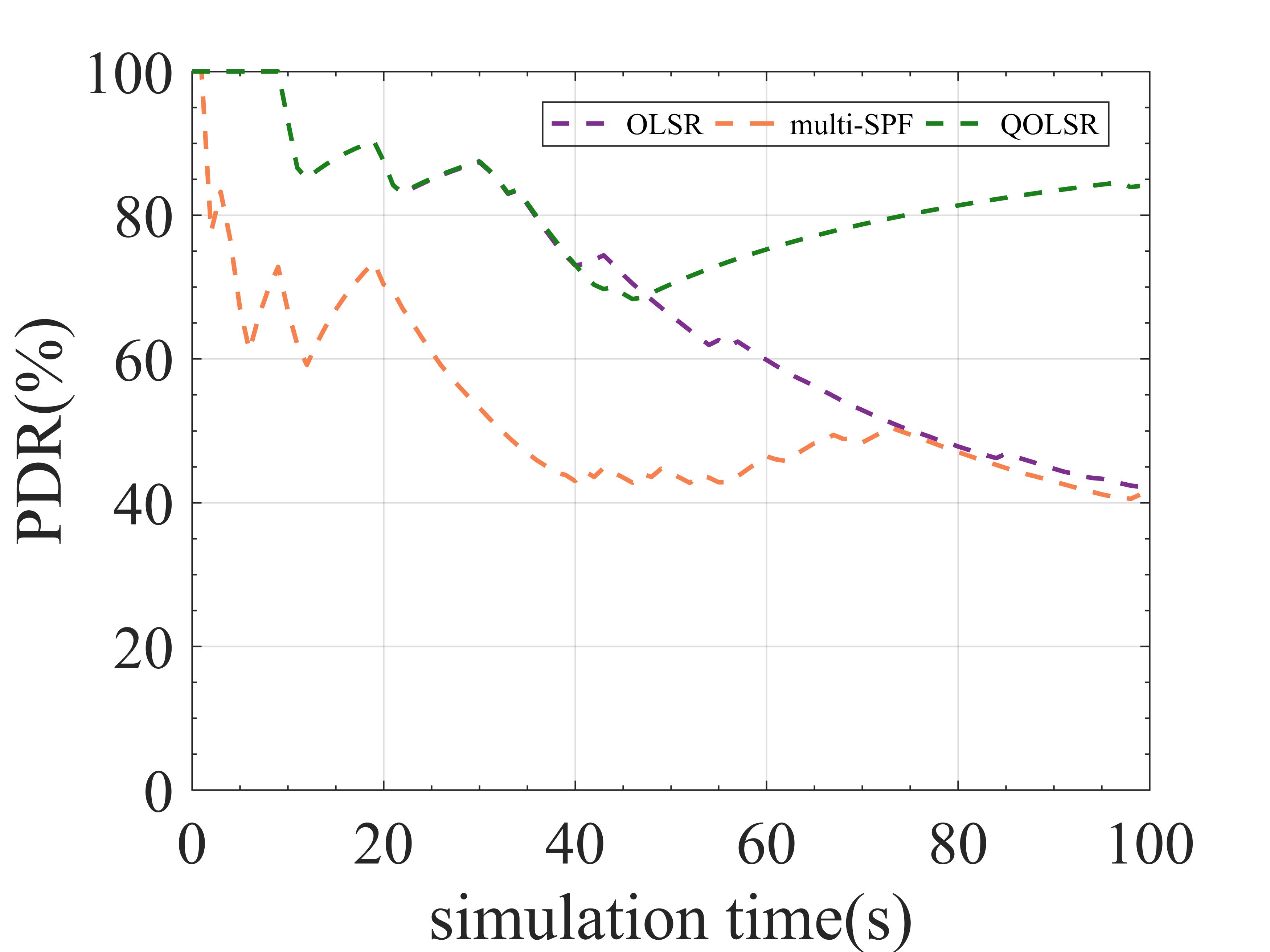}}
	\caption{Comparison of OLSR, multi-SPF, and~QOLSR in real-time~PDR. (\textbf{a}) Communication \mbox{level = 0.2}. (\textbf{b}) Communication level = 0.4. (\textbf{c}) Communication level = 0.6. (\textbf{d}) Communication \mbox{level = 0.8}.}
	\label{fig9}
\end{figure}

\begin{figure}[H]
	\centering
	\includegraphics[width=10.5 cm]{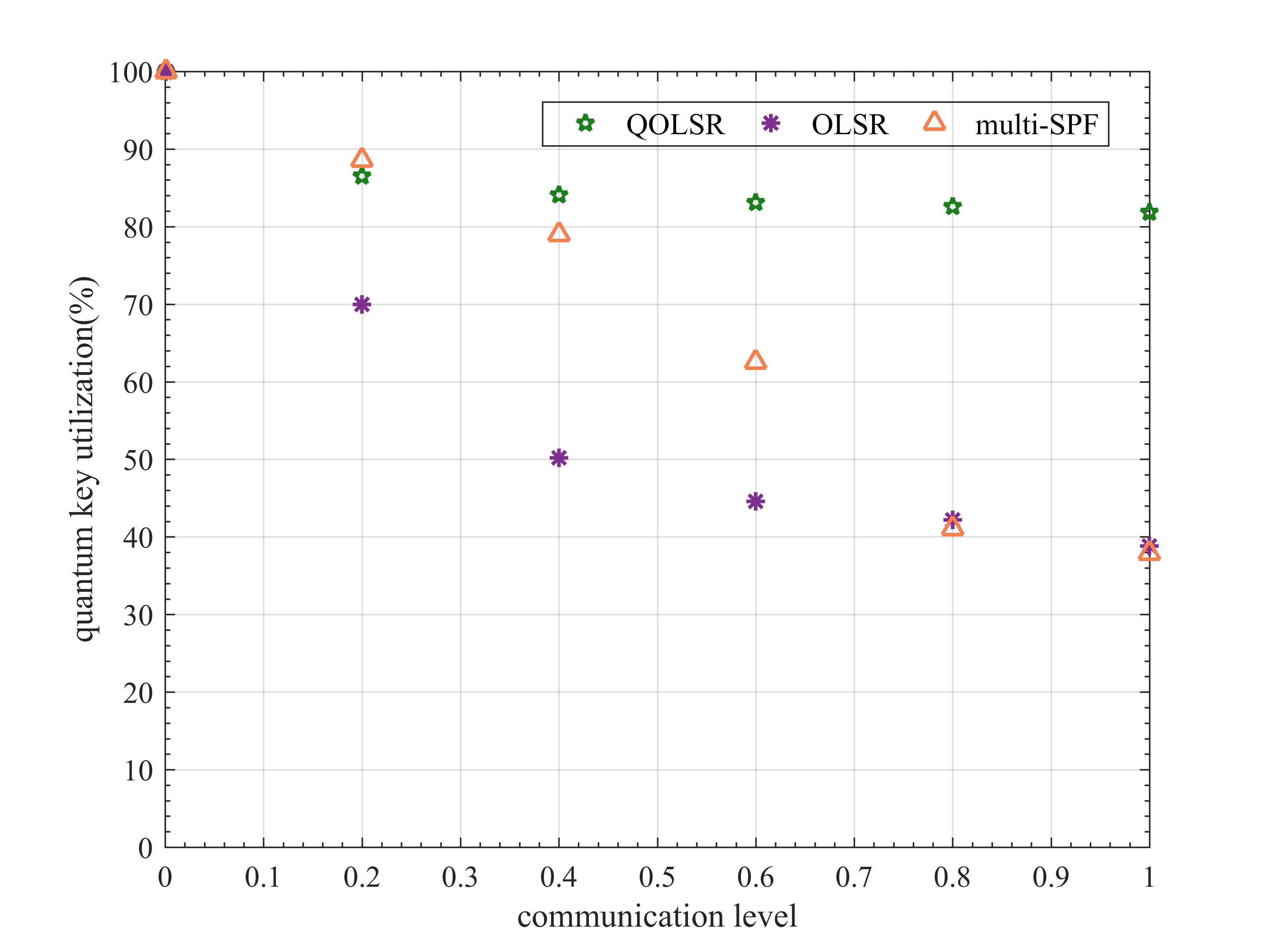}
	\caption{Comparison of OLSR, multi-SPF and QOLSR in quantum secure key utilization\label{fig10}}
\end{figure}

When the communication level is low, the~performance of QOLSR and multi-SPF is close and is higher than that of OLSR, as~shown in Figure~\ref{fig9}. As~communication level increases, the~PDRs of OLSR and multi-SPF decrease sharply. However, the~PDR of QOLSR still remains high after fluctuation. This is mainly due to link-state awareness and routing based on key recovery capability. Even if the network’s communication demand is enormous, it can still route packets in a reasonable manner, and~the network will not be paralyzed. It can be seen from the experimental result statistics in Figure~\ref{fig10} that the efficient routing protocol proposed here has very high quantum secure key utilization at any communication level, which is significantly higher than the original, unimproved routing protocol. In~particular, it can be seen that when the communication level is 1, the~quantum secure key utilization rate of the original OLSR is 38.81$\%$, while the quantum secure key utilization rate of the QOLSR we proposed is 81.84$\%$, which is 2.1 times higher than that of the original OLSR. This is mainly due to improved link-state awareness, which reduces the number of packets lost during path switching, and~improved path optimization based on key recovery capability, which reduces the path switching frequency and improves quantum secure key utilization. At~low communication levels, the~multi-SPF quantum secure key functions effectively. However, its performance decreases rapidly when the communication level rises. This is primarily due to multi-path SPF’s selection approach, which is based on the amount of the key pool's remaining keys. When the communication level is low, the~key consumption rate is low. Therefore, the~key consumption rate has a minimal impact on communication. However, when the amount of communication increases, so does the key consumption rate. For~the performance of multi-SPF, which does not take into account key generation, the~consumption rate in routing is reduced dramatically. In~particular, when the communication level is 0.01 and 0.2, the~quantum key utilization of multi-SPF is slightly higher than that of QOLSR. This is mainly due to the low frequency of path switching at extremely low communication levels. The~link pre-judgment in the link-state awareness mechanism of QOLSR may cause path switching to occur earlier than multi-SPF. As~a result, the~switching frequency of QOLSR is higher than that of multi-SPF. Moreover, due to the low key consumption rate, the~advantage of choosing a path with a long sustainable working time based on key recovery capability cannot be exerted. Consequently, the~quantum key utilization rate of QOLSR is slightly lower than that of multi-SPF when the communication level is extremely low. However, with~increased communication, the~quantum key utilization of QOLSR is significantly higher than that of multi-SPF. The~performance of QOLSR, on~the other hand, is relatively stable even at high communication levels. This means the quantum secure key utilization of the QKD network with the optimized routing protocol is significantly~improved.

\unskip

\unskip

\begin{figure}[H]
	\centering
	\subfloat[\centering]{
		\label{fig11-1}
		\includegraphics[scale=0.04]{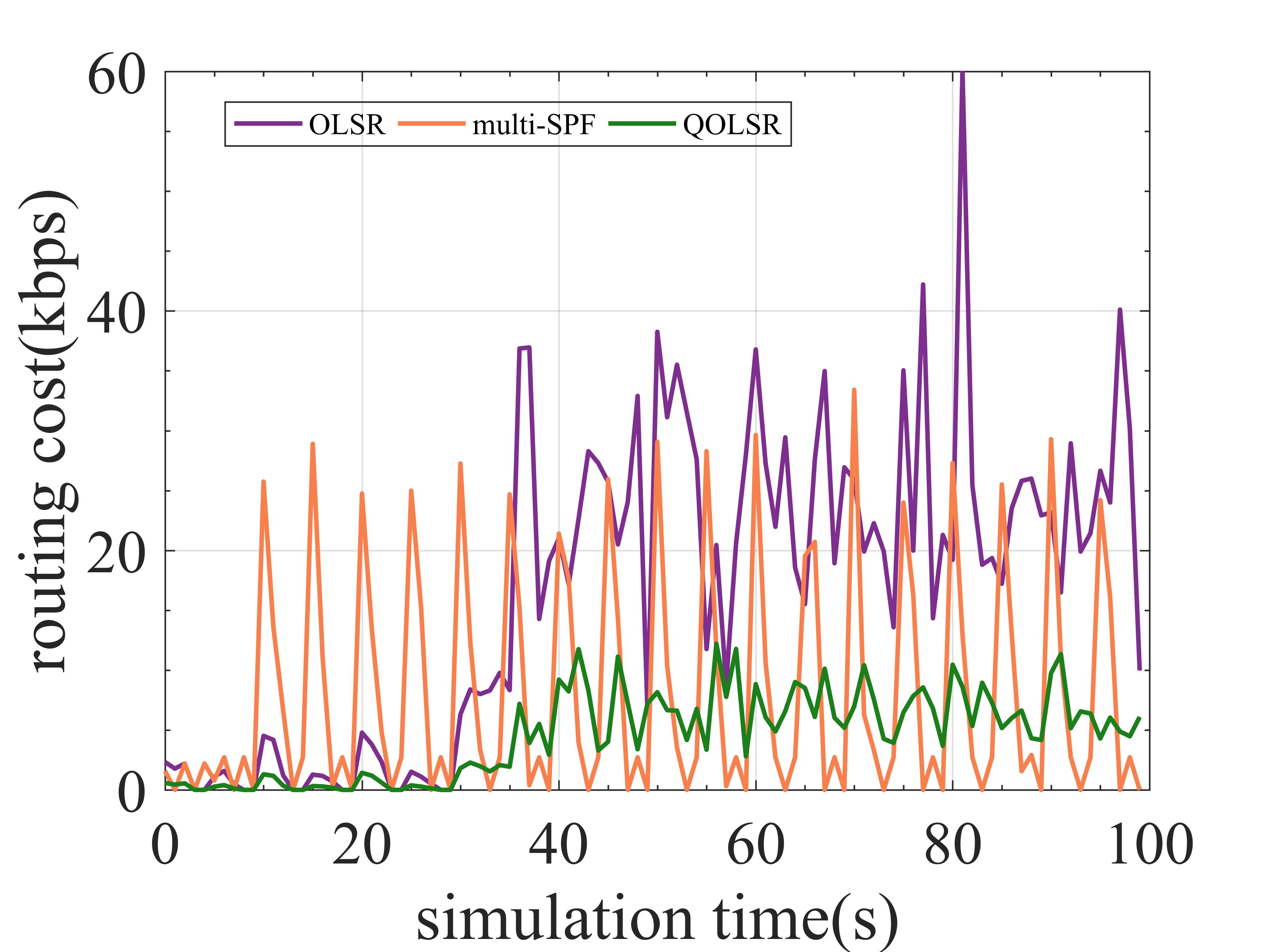}}
	\subfloat[\centering]{
		\label{fig11-2}
		\includegraphics[scale=0.04]{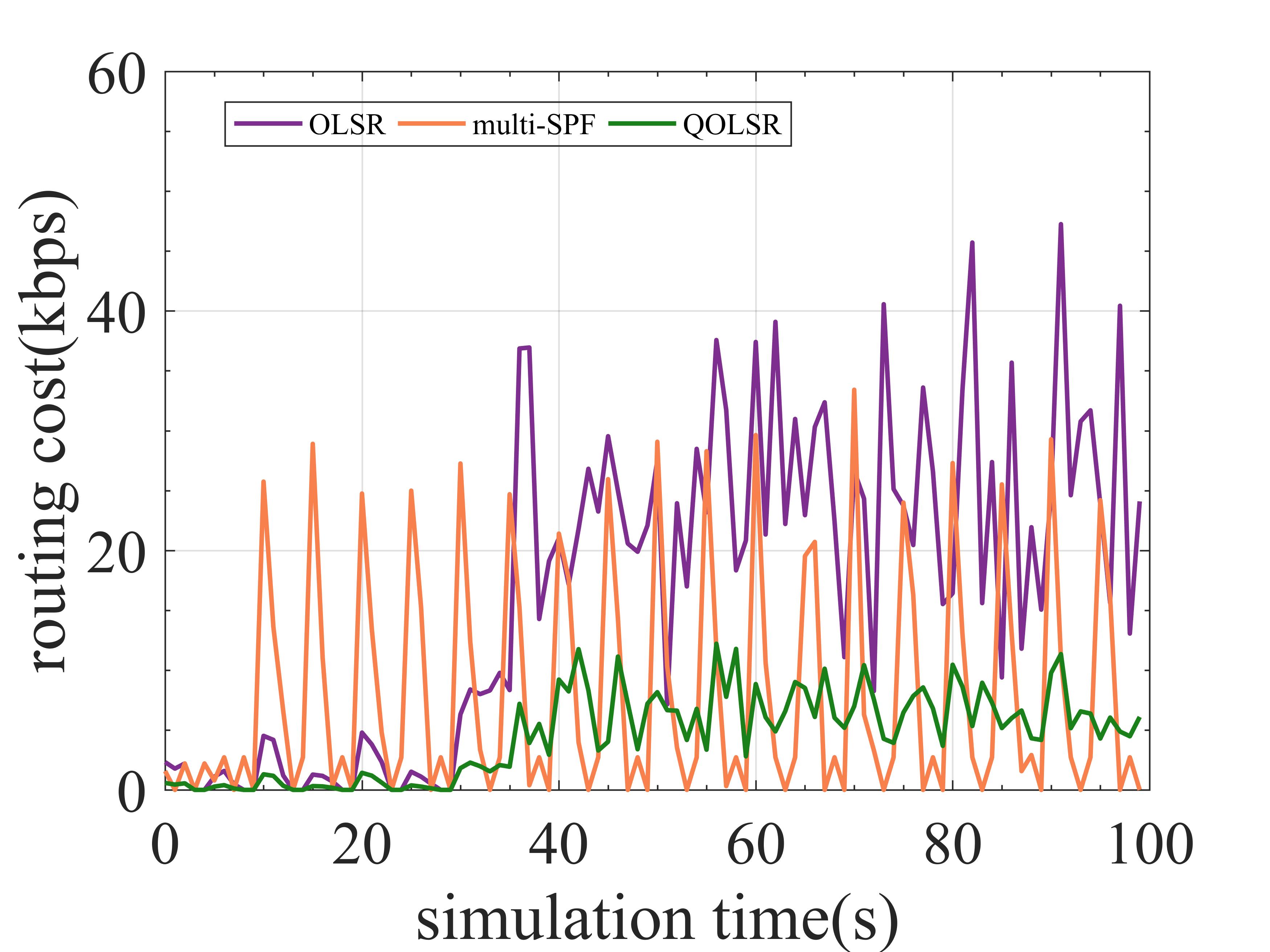}}\\ 
	\subfloat[\centering]{
		\label{fig11-3}
		\includegraphics[scale=0.04]{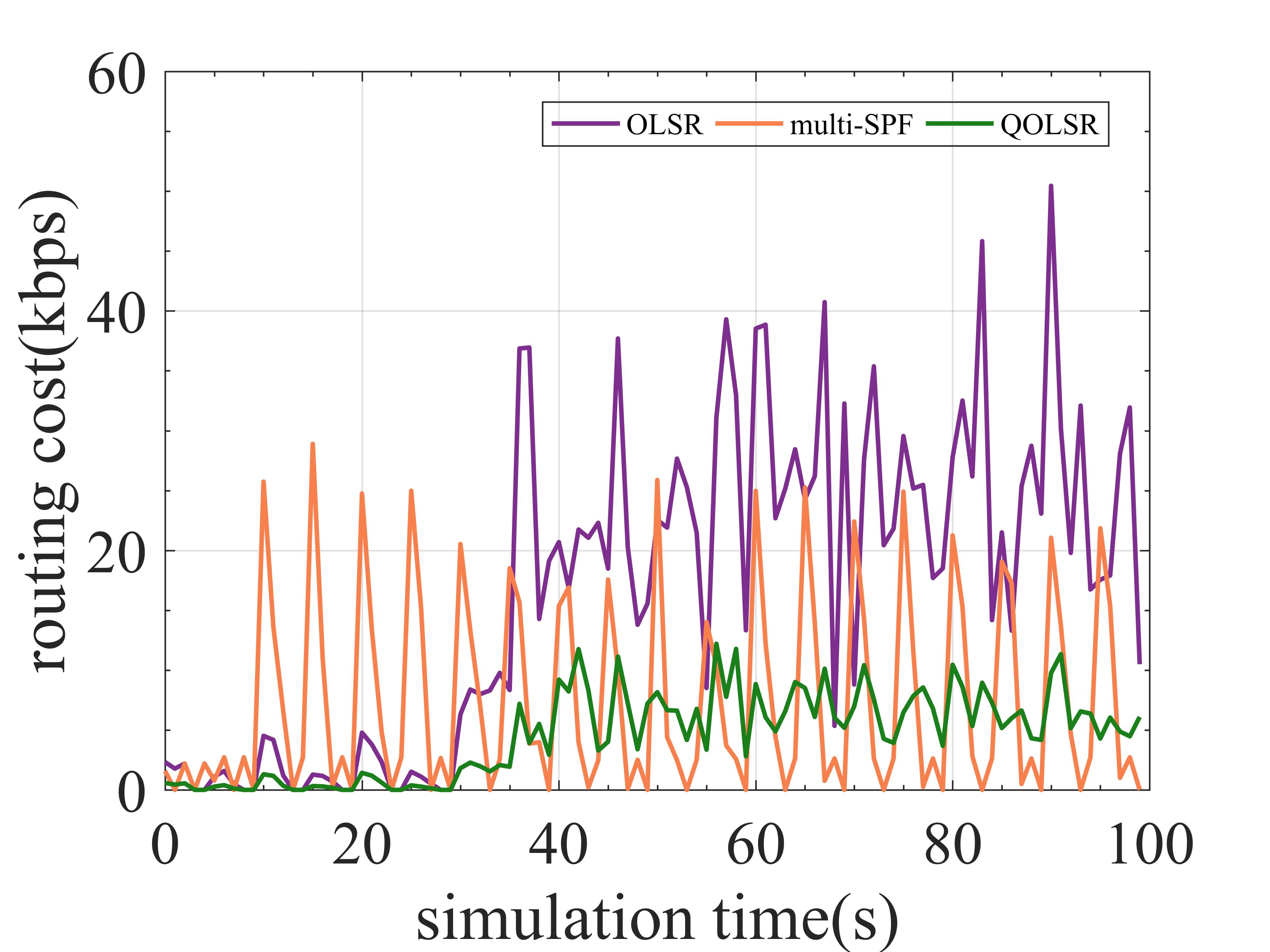}}
	\subfloat[\centering]{
		\label{fig11-4}
		\includegraphics[scale=0.04]{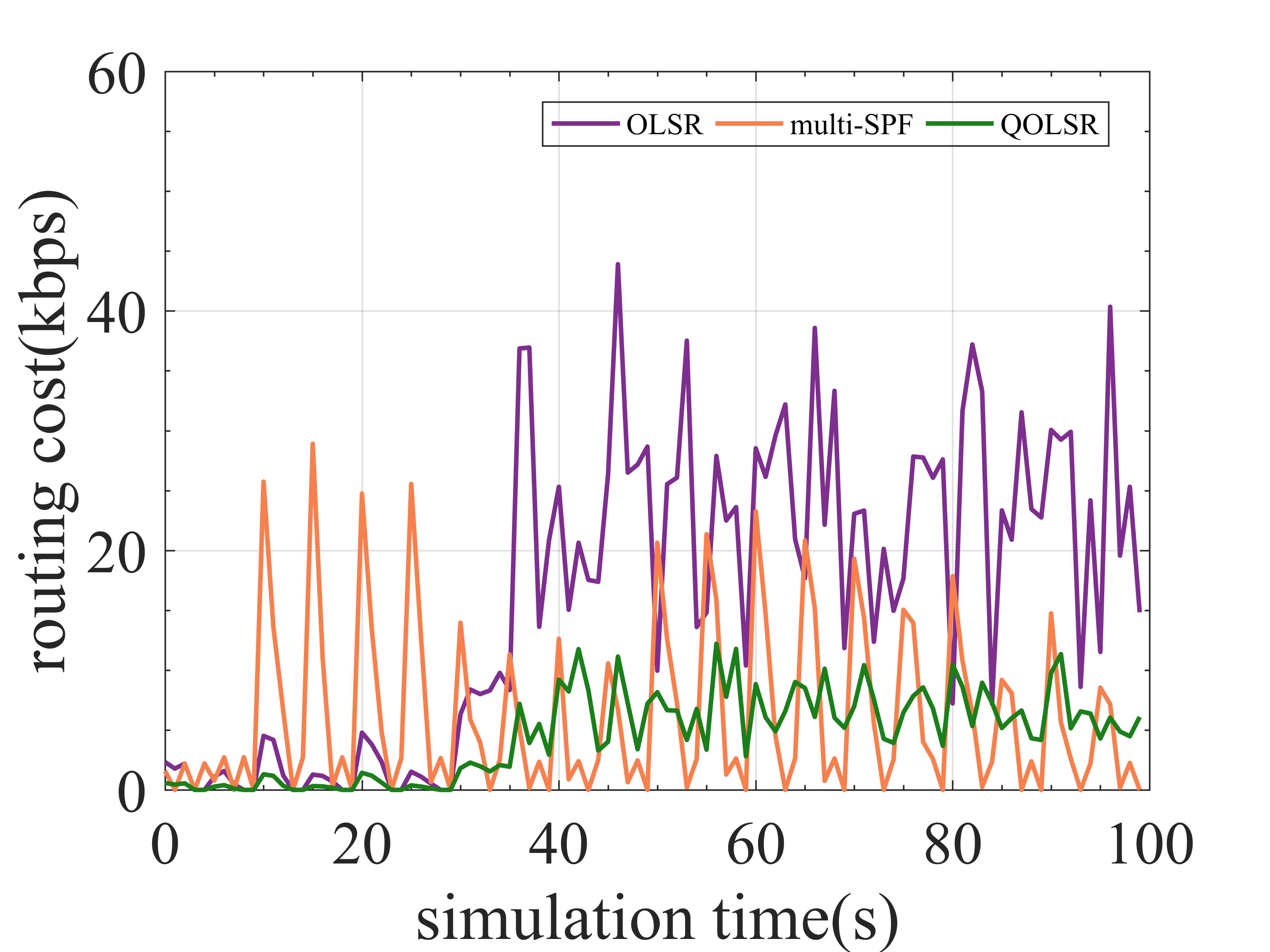}}
	\caption{Comparison of OLSR, multi-SPF, and~QOLSR in real-time routing~cost. (\textbf{a}) Communication \mbox{level = 0.2}. (\textbf{b}) Communication level = 0.4. (\textbf{c}) Communication level = 0.6. (\textbf{d}) Communication \mbox{level = 0.8}.}
	\label{fig11}
\end{figure}

\begin{figure}[H]
	\centering
	\includegraphics[width=10.5 cm]{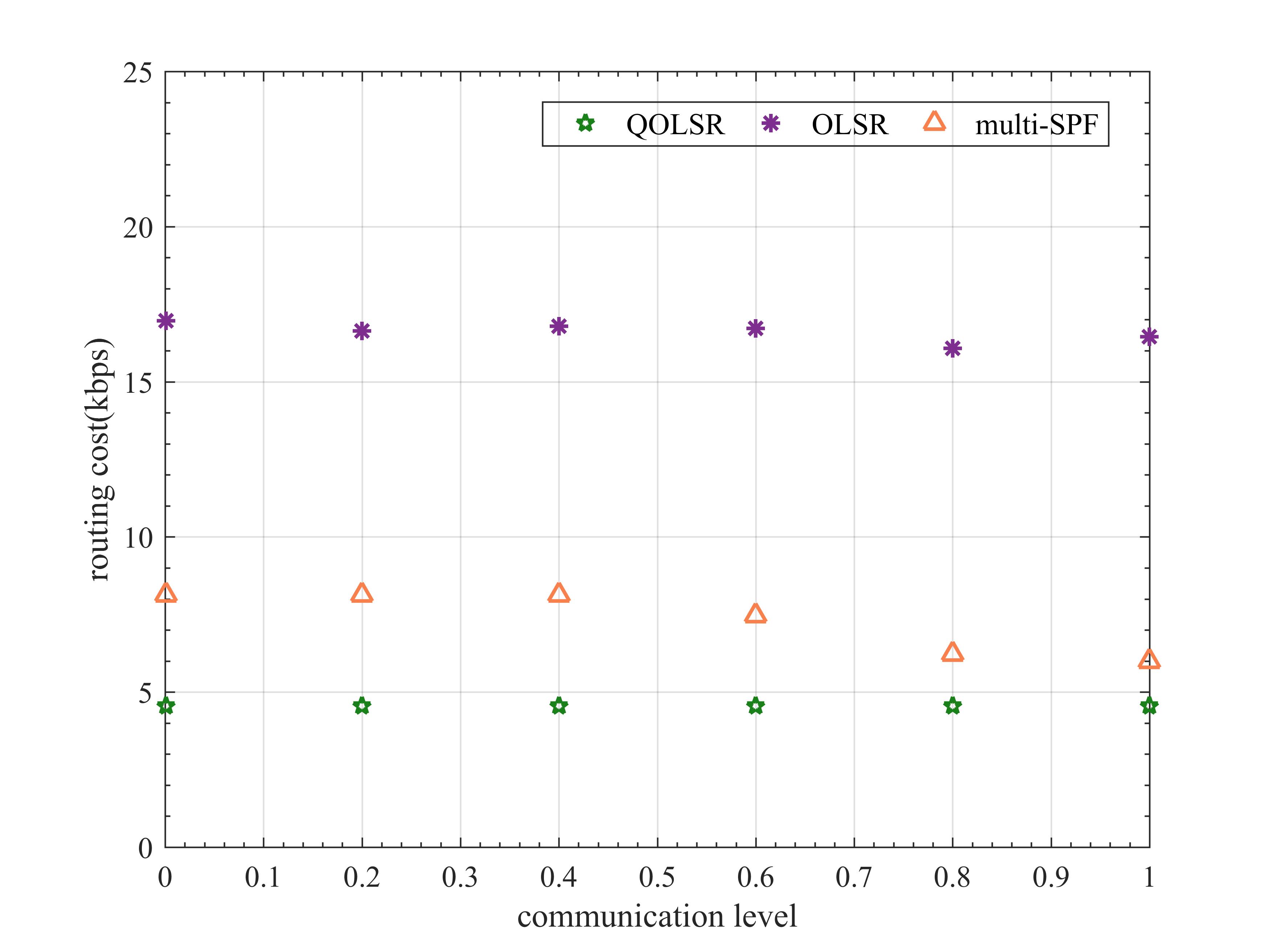}
	\caption{Comparison of OLSR, multi-SPF and QOLSR in routing cost\label{fig12}}
\end{figure}

\begin{figure}[H]
	\centering
	\subfloat[\centering]{
		\label{fig13-1}
		\includegraphics[scale=0.04]{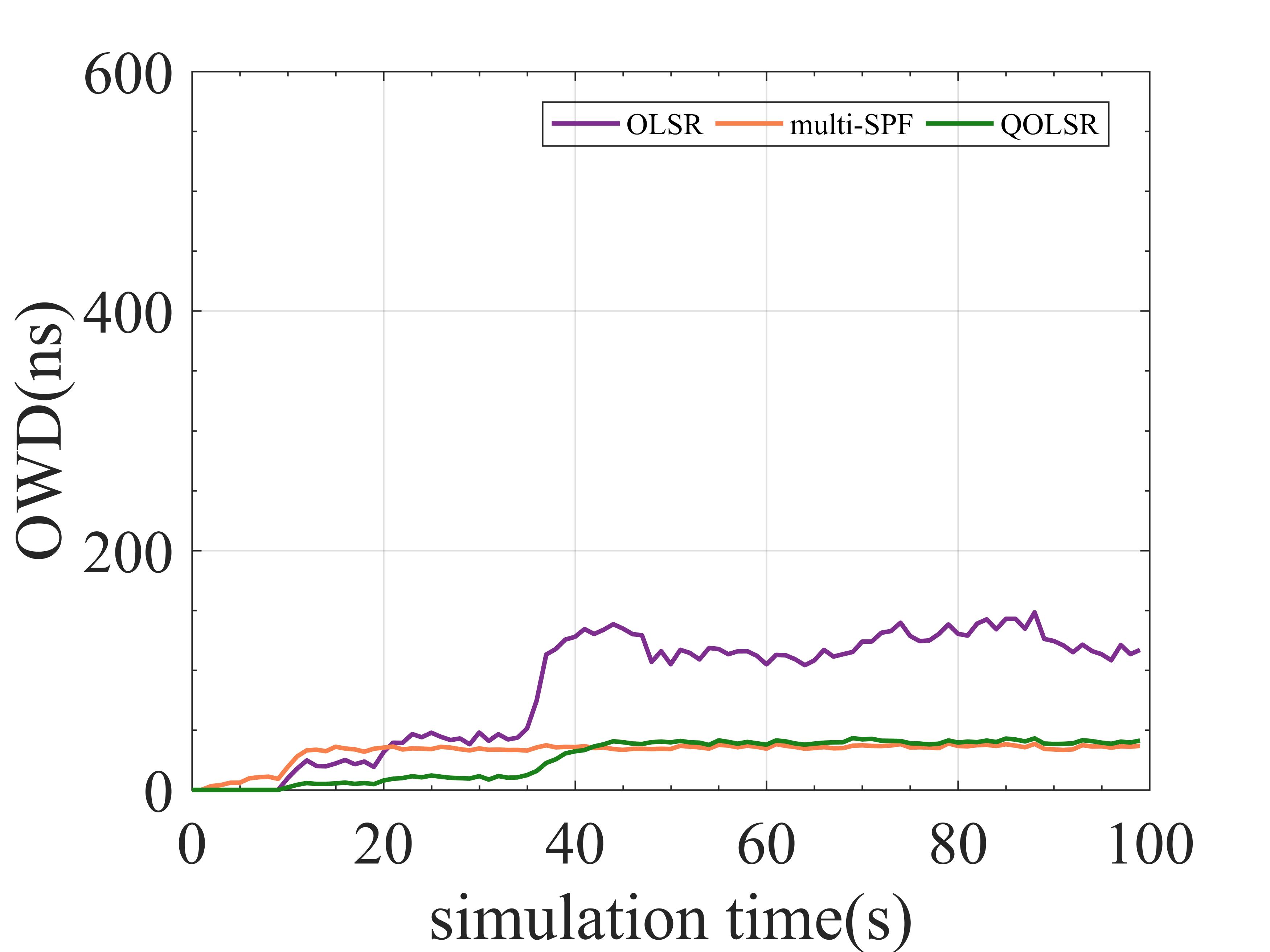}}
	\subfloat[\centering]{
		\label{fig13-2}
		\includegraphics[scale=0.04]{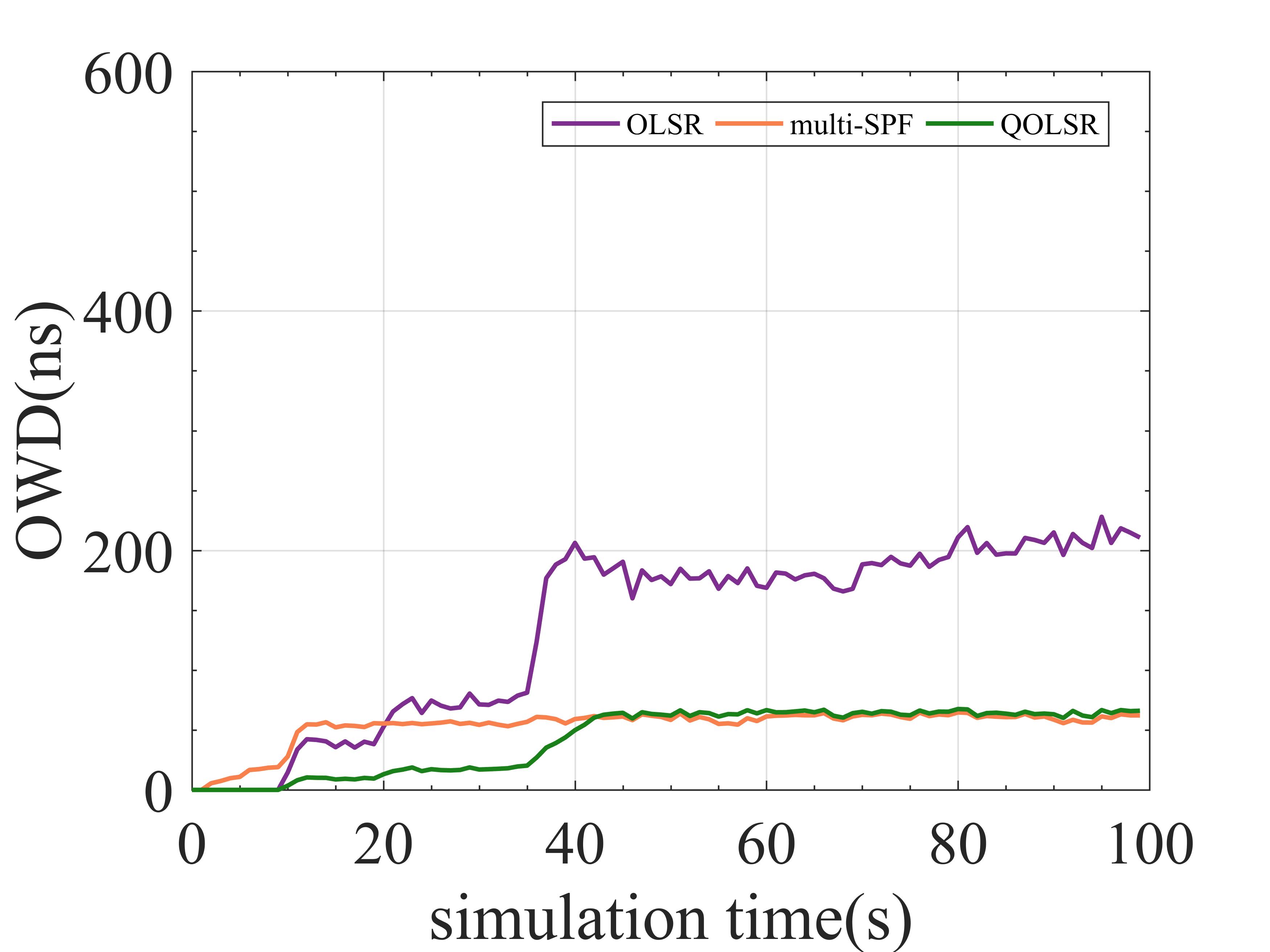}}\\ 
	\subfloat[\centering]{
		\label{fig13-3}
		\includegraphics[scale=0.04]{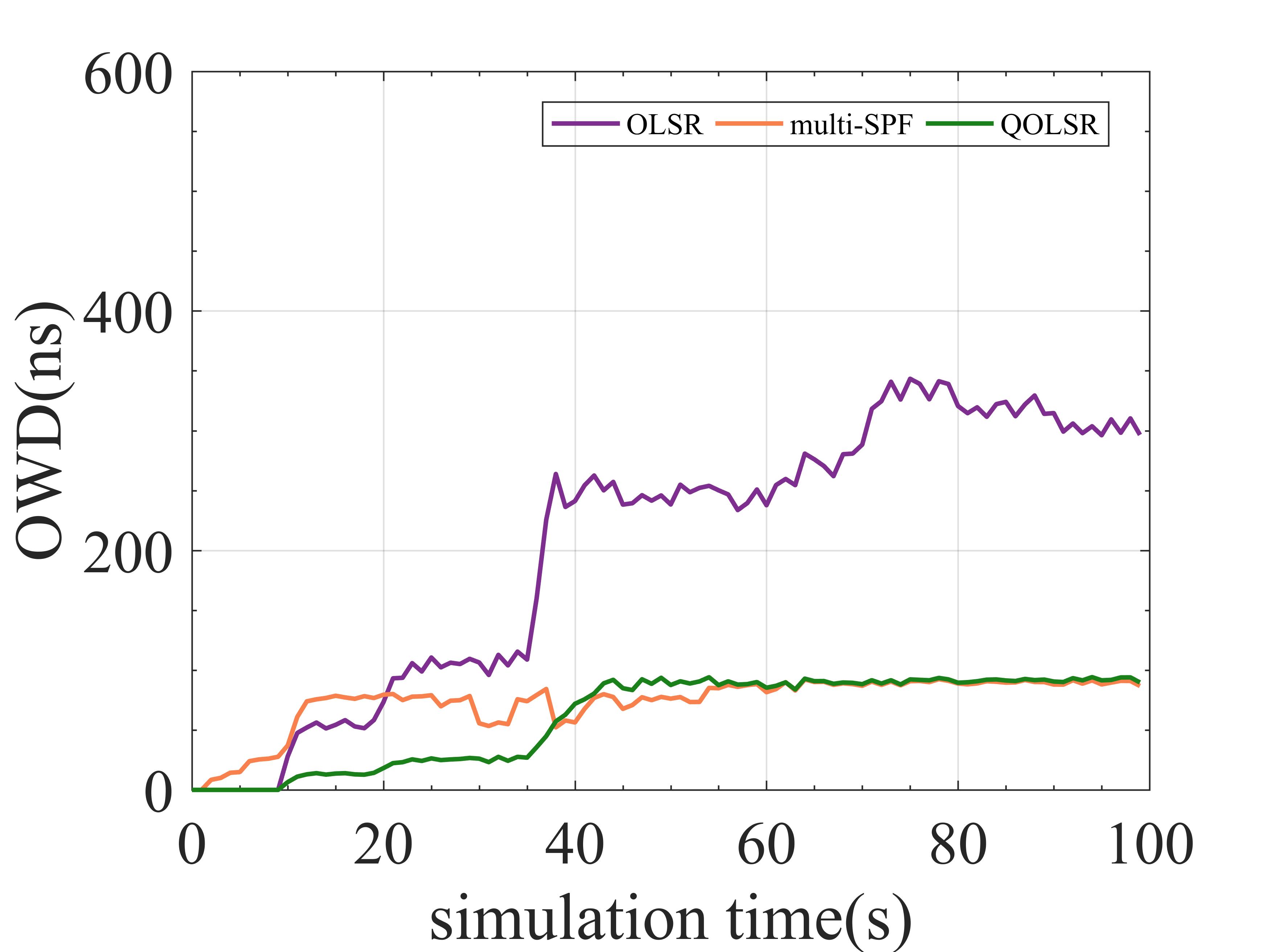}}
	\subfloat[\centering]{
		\label{fig13-4}
		\includegraphics[scale=0.04]{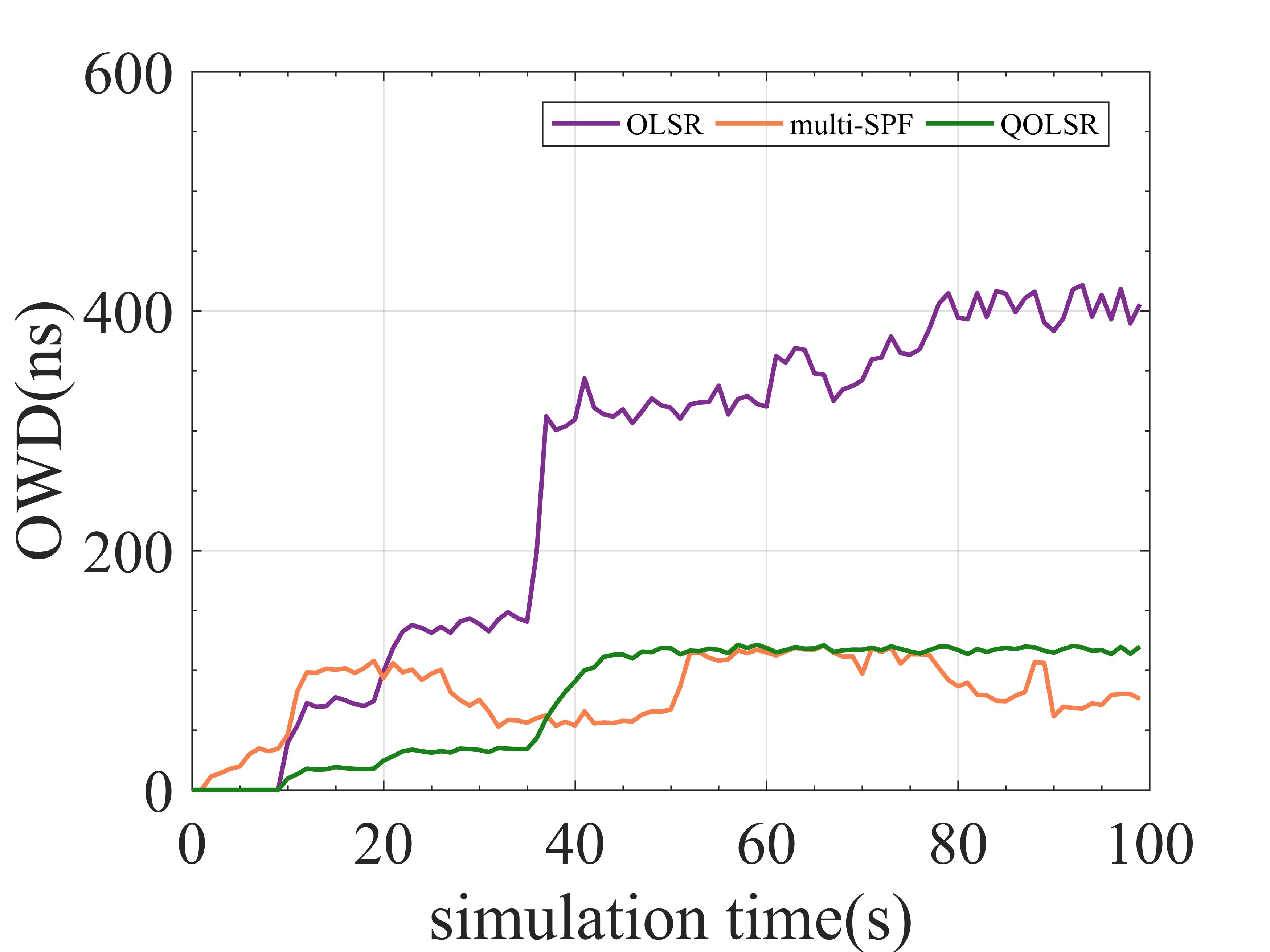}}
	\caption{Comparison of OLSR, multi-SPF, and~QOLSR in real-time one-way delay (OWD). (\textbf{a})~Communication \mbox{level = 0.2}. (\textbf{b}) Communication level = 0.4. (\textbf{c}) Communication level = 0.6. (\textbf{d})~Communication \mbox{level = 0.8}.}
	\label{fig13}
\end{figure}

\begin{figure}[H]
	\centering
	\includegraphics[width=10.5 cm]{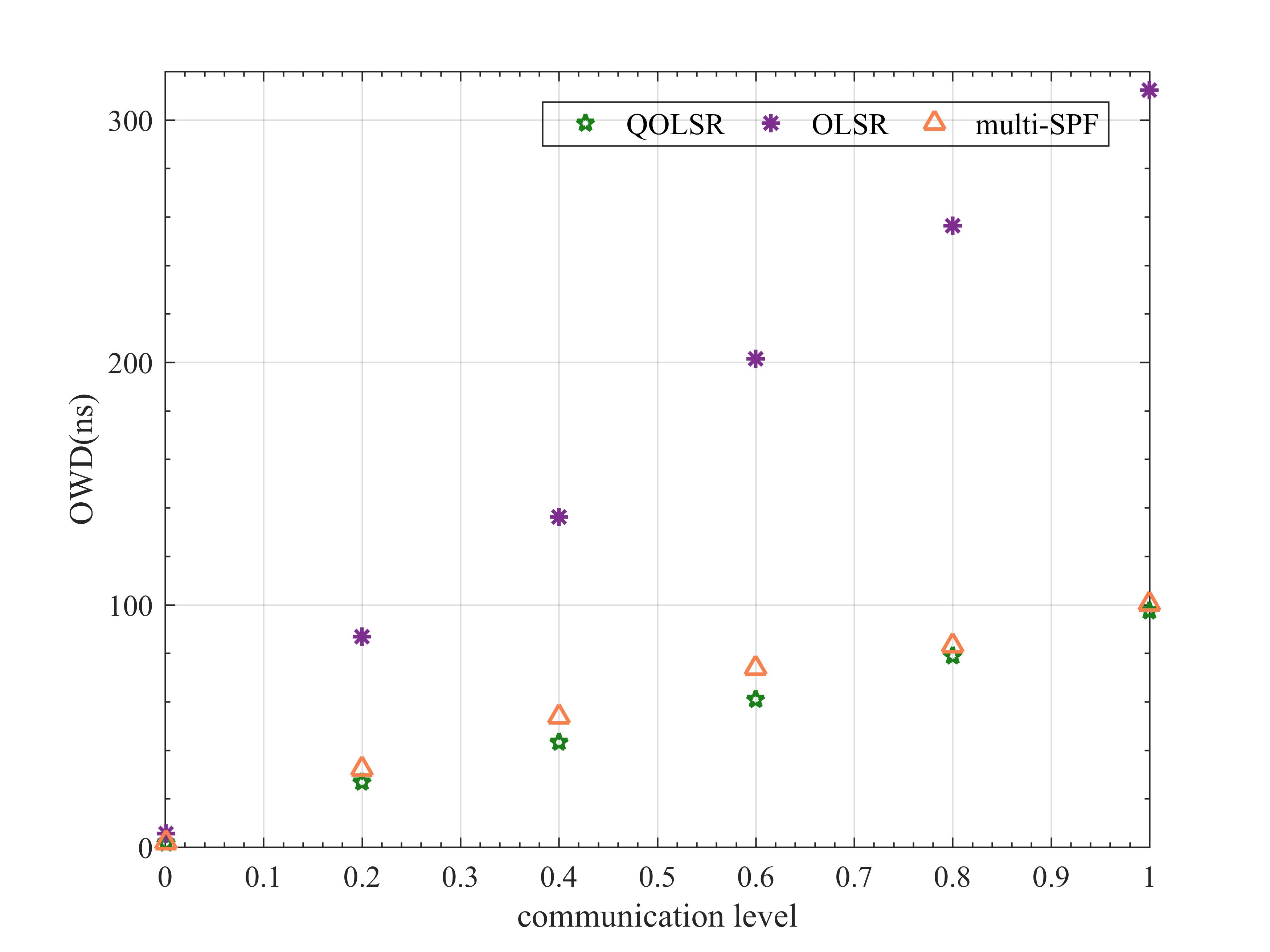}
	\caption{Comparison of OLSR, multi-SPF and QOLSR in one-way delay (OWD)\label{fig14}}
\end{figure}

In addition, since routing packets also need to be encrypted and protected, the~routing cost and one-way delay (OWD) time are also compared for a more comprehensive evaluation of the QOLSR protocol’s performance. The~number of quantum keys consumed by routing packets in communication is referred to as routing cost. The~delay of data packets from transmitting to receiving is referred to as OWD. Although~routing cost fluctuates during communication, the~routing cost of QOLSR is always minimal, as~shown in Figure~\ref{fig11}. The~routing cost of QOLSR is substantially lower than that of original OLSR and multi-SPF, as~seen in Figure~\ref{fig12}. As~a result of the increased link-state awareness and path optimization suggested in this research, the~quantum secure keys spent by routing packets is lowered. This cuts down on the waste of quantum secure keys. Herein, it can be seen from Figure~\ref{fig13} that the delay of QOLSR during communication is similar to that of multi-SPF, and~it has been kept at a low level. The~OWD of QOLSR is much lower than that of original OLSR, as~illustrated in Figure~\ref{fig14}, which further demonstrates that the quality of service in QKD networks with QOLSR has~improved.

\unskip

\unskip

\unskip

\unskip

In summary, the~simulation results show that QOLSR not only has better quantum secure key utilization, it also performs well in routing cost and OWD. This is statistically significant because in QKD networks, quantum secure key resources are very precious. A~key reason for this is that the efficient link-state awareness mechanism of the routing protocol is based on the amount of remaining key resources, which reserves time for path switching and reduces packet loss. Moreover, efficient path optimization is based on key recovery capability, which reduces the frequency of path switching. As~a result, the~efficiency of the proposed QOLSR routing protocol is~verified.

\section{Conclusions}\label{sec4}
This paper has presented an efficient routing protocol, QOLSR, for~QKD networks. To~be exact, in~this routing protocol, we primarily built efficient link-state awareness and path optimization based on key recovery capability. Simulations demonstrated that QOLSR has significant performance improvements. As~a result, we believe that the proposed routing algorithm has the potential to dramatically improve QKD networks' quantum secure key~utilization.

\end{document}